\documentclass[prl,twocolumn,amsmath,amssymb,nofootinbib]{revtex4-1}
\usepackage{graphicx}
\usepackage{amsmath}
\usepackage{MnSymbol}
\usepackage{mathrsfs}
\usepackage{colortbl}
\usepackage[table]{xcolor}
\usepackage{tabularx}
\usepackage{amssymb}

\makeatletter

\newenvironment{FIG.here}
  {\def\@captype{FIG.}}
  {}
\makeatother

\begin{document}
\title{Robust Chiral Optical Force via Electric Dipole Interactions, Inspired by a Sea Creature}

\author{Robert P. Cameron}
\email{robert.p.cameron@strath.ac.uk}
\affiliation{SUPA and Department of Physics, University of Strathclyde, Glasgow G4 0NG, United Kingdom}

\author{Duncan McArthur}
\affiliation{SUPA and Department of Physics, University of Strathclyde, Glasgow G4 0NG, United Kingdom}

\author{Alison M. Yao}
\affiliation{SUPA and Department of Physics, University of Strathclyde, Glasgow G4 0NG, United Kingdom}

\author{Nick Vogeley}
\affiliation{Institute of Applied Physics, University of Bonn, 53115 Bonn, Germany}

\author{Daqing Wang}
\affiliation{Institute of Applied Physics, University of Bonn, 53115 Bonn, Germany}

\begin{abstract}
Inspired by a sea creature, we identify a robust chiral optical force that \textcolor{black}{can be several orders of magnitude stronger than others proposed to date, pushing} the opposite enantiomers of \textcolor{black}{orientated} chiral molecules towards regions of orthogonal linear \textcolor{black}{rather than} circular polarization in an optical field via electric dipole interactions. Our chiral optical force applies \textcolor{black}{off resonance} to essentially all chiral molecules, including isotopically chiral \textcolor{black}{species} which are notoriously difficult to \textcolor{black}{resolve} using existing methods. \textcolor{black}{We conclude by proposing a realistic experiment supported by numerical simulations. This work could enable the first demonstration of a chiral optical force for individual chiral molecules and thus the first non-destructive enantioseparation using light.}
\end{abstract} 

\date{\today}
\maketitle

Chiral optical forces have now been demonstrated in the laboratory for relatively large objects including chiral liquid crystal microspheres \cite{Cipparrone11a, Hernandez13a, Tkachenko13a, Donato14a, Tkachenko14a, Tkachenko14b, Kravets19a, Kravets19b}, chiral cantilevers \cite{Zhao17a}, chiral nanoparticles \textcolor{black}{\cite{Yamanishi22a, Tkachenko26a}}, and chiral nanostructures \cite{Yamanishi23a}. The realization of such forces at the \textit{molecular} scale remains an outstanding challenge, however; nobody has demonstrated a chiral optical force for individual chiral molecules, in spite of potential applications including the separation of opposite enantiomers \cite{Canaguier13a, Cameron14b, Cameron14c, Bradshaw15a, Hayat15a, Jones17a, Forbes22a, Martinez-Romeu24a}, tests of fundamental physics \cite{Stickler21a}, the distillation of novel forms of matter \cite{Isaule22a}, and more \cite{Marichez19a, Kakkanattu21a, Genet22a, Golat24a, Habibovic24a, Toftul24a}.

The main difficulty is that the chiral optical forces proposed to date for chiral molecules are extremely weak. One representative example is the use of an optical helicity lattice to push opposite enantiomers towards regions of opposite \textit{circular} polarization via interference between electric dipole and magnetic dipole interactions \cite{Cameron14b, Cameron14c}, giving a chiral optical force of the form
\begin{align}
\mathbf{F}^\prime={}&-\frac{\omega}{\epsilon_0 c}G^\prime(\omega)\pmb{\nabla}_{\mathbf{R}_0}h, \label{Fh}
\end{align}
where $G^\prime(\omega)$ is the molecule's electric dipole-magnetic dipole polarizability pseudoscalar and \textcolor{black}{$\pmb{\nabla}_{\mathbf{R}_0}h$ is the gradient of the lattice's helicity density, $\omega$ being the lattice's angular frequency and $\mathbf{R}_0=Z_0\hat{\mathbf{z}}+Y_0\hat{\mathbf{y}}+X_0\hat{\mathbf{x}}$ being the molecule's position}; $\mathbf{F}^\prime$ is typically smaller than traditional optical gradient forces by three \textcolor{black}{to five orders of magnitude}. Note, moreover, that Eq. \ref{Fh} neglects the molecule's anisotropy, severely limiting its applicability to the molecule's unperturbed rotational ground state.

In this Letter, we identify a fundamentally different chiral optical force $\mathbf{F}\propto\alpha_{yx}(\omega)$ \textcolor{black}{that can be several orders of magnitude stronger than the optical helicity gradient force $\mathbf{F}^\prime\propto G^\prime(\omega)$, pushing the opposite enantiomers of a chiral molecule towards regions of orthogonal \textit{linear} rather than circular polarization in an optical field via electric dipole interactions alone; $\alpha_{yx}(\omega)$ is a component of the molecule's electric dipole-electric dipole polarizability tensor.} The key ingredient is suitable molecular \textit{orientation}, explicitly realized in this Letter using a combination of static and traveling-wave fields. Our theory fully accounts for the molecule's anisotropy and rotational degrees of freedom (usually neglected in the molecule-light interactions \cite{Canaguier13a, Cameron14b, Cameron14c, Bradshaw15a, Hayat15a, Jones17a, Forbes22a, Martinez-Romeu24a, Stickler21a, Isaule22a, Marichez19a, Kakkanattu21a, Genet22a, Golat24a, Habibovic24a, Toftul24a}) and we use quantum chemical predictions of molecular properties rather than inflated hypothetical values. \textcolor{black}{We conclude by proposing a realistic experiment supported by numerical simulations.}

\begin{figure}[h!]
\centering
\includegraphics[width=\linewidth]{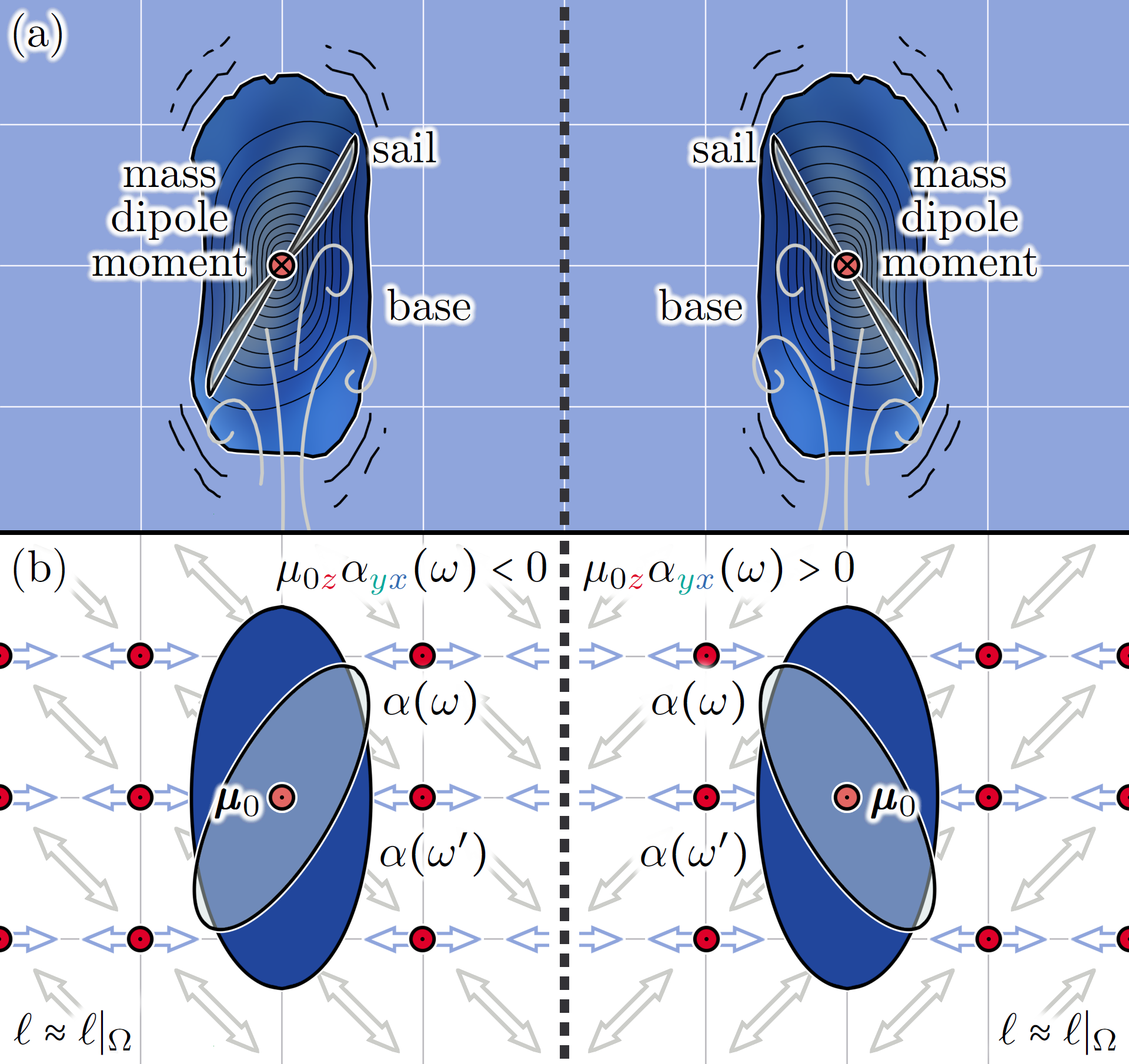}
\caption{\small (a) The key elements of \textit{Velella velella} \textcolor{black}{with the specimen's orientation chosen to help illustrate our analogy}. (b) The analogous molecular properties with the molecule's orientation assumed close to a minimum of the rotational potential energy $V$. \textcolor{black}{The red arrows indicate the direction of the static field $\mathbf{E}_0$, the double-headed blue arrows indicate the polarization of the traveling-wave field $\tilde{\mathbf{E}}^\prime$, and} the double-headed grey arrows indicate the orthogonal linear polarizations of the standing-wave field $\tilde{\mathbf{E}}$ that the molecules are pushed towards by our chiral optical force $\mathbf{F}$.}
\label{Fig1}
\end{figure}

The inspiration for our chiral optical force is \textit{Velella velella}, the ``by-the-wind sailor''; a hydroid colony resembling a raft. \textcolor{black}{The mass dipole moment\footnote{The specimen's centres of gravity and buoyancy can be thought of as spatially separated masses of equal magnitude and opposite sign, forming a mass dipole moment.}, base, and sail form a chiral geometry, giving two distinct mirror-image forms as illustrated in Fig. \ref{Fig1}(a). Together, gravity and the water help give a specimen preferred orientations via the mass dipole moment and base such that the two forms are sometimes separated by the wind \cite{Mackie62a, Neville76a, Iwanicki25a}.} We construct our chiral optical force by analogy \textcolor{black}{as follows}. Instead of a \textit{Velella velella} specimen, we consider a polar, diamagnetic, chiral molecule in its vibronic ground state, located at the origin say. \textcolor{black}{We find in general for such molecules that the permanent electric dipole moment $\pmb{\mu}_0$, polarizability tensor $\alpha(\omega^\prime)$ at a far off-resonance angular frequency $\omega^\prime$, and polarizability tensor $\alpha(\omega)$ at a different far off-resonance angular frequency $\omega$ form a chiral geometry comparable to \textit{Velella velella} as illustrated in Fig. \ref{Fig1}(b)}. Instead of gravity, we consider a strong and homogeneous static electric field
\begin{align}
\mathbf{E}_0={}&\mathcal{E}_z\hat{\mathbf{z}}, \label{StaticE}
\end{align}
where $\mathcal{E}_z$ determines the field's strength and direction; $\mathbf{E}_0$ partially orientates the molecule (pitch and roll) via a dc Stark interaction with $\pmb{\mu}_0$. Instead of the water, we consider an intense and far off-resonance linearly polarized traveling wave with complex electric field 
\begin{align}
\tilde{\mathbf{E}}^\prime={}&\mathcal{E}_x^\prime\mathrm{e}^{\mathrm{i}(k^\prime y-\omega^\prime t)}\hat{\mathbf{x}}, \label{ComplexEPrime}
\end{align}
where $\mathcal{E}_x^\prime$ determines the \textcolor{black}{wave's} amplitude and overall sign and $k^\prime=\omega^\prime/c$ \textcolor{black}{is} the \textcolor{black}{wave's} angular wavenumber; $\tilde{\mathbf{E}}^\prime$ aligns the molecule (yaw) via an ac Stark interaction with $\alpha(\omega^\prime)$. \textcolor{black}{Together, $\mathbf{E}_0$ and $\tilde{\mathbf{E}}^\prime$ orientate the molecule \cite{Friedrich99a, Sakai03a, Minemoto03a, Tanji05a} such that the expectation value of the polarizability component $\alpha_{yx}(\omega)$ has opposite signs for opposite enantiomers.} Instead of the wind, we consider a relatively weak and far off-resonance lin$\perp$lin standing wave with complex electric field
\begin{align}
\tilde{\mathbf{E}}={}&\mathcal{E}_y\mathrm{e}^{\mathrm{i}(kz-\omega t)}\hat{\mathbf{y}}+\mathrm{i}\mathcal{E}_x\mathrm{e}^{\mathrm{i}(-kz-\omega t)}\hat{\mathbf{x}}, \label{ComplexE}
\end{align}
where $\mathcal{E}_y$ determines the amplitude and overall sign of the $y$ polarized wave, $\mathcal{E}_x$ determines the amplitude and overall sign of the $x$ polarized wave, and $k=\omega/c$ \textcolor{black}{is} the angular wavenumber \textcolor{black}{of the waves}; our chiral optical force $\mathbf{F}$ pushes opposite enantiomers towards regions of orthogonal linear polarization in $\mathbf{E}$ via a spatially varying ac Stark interaction with $\alpha_{yx}(\omega)$.

\textcolor{black}{Our introduction of the travelling-wave field $\tilde{\mathbf{E}}^\prime$ in this Letter is a vital improvement over our previous work \cite{Cameron23a} that opens the door to the strong-field regime and thus a robust chiral optical force with excellent overall enantioselectivity. In \cite{Cameron23a}, we instead assumed quantization along the $x$ axis without explicit justification and restricted our attention to the weak-field regime, giving chiral optical forces orders of magnitude smaller and with much poorer overall enantioselectivity than those predicted in this Letter.}

\textcolor{black}{A distinction between this work and proposals for resolution using enantiomer-specific state transfer \cite{Eibenberger17a, Eibenberger-Arias24a, Ordonez25a} is that our chiral optical force $\mathbf{F}$ is based solely on off-resonance interactions, meaning it does not require a specific energy-level structure and a single field configuration could work for many different molecular species.} The precise values of the angular frequencies $\omega^\prime$ and $\omega$ are not critical, however $\omega^\prime$ and $\omega$ must \textcolor{black}{differ} to help ensure that the principal axes of the molecule's $\alpha(\omega^\prime)$ and $\alpha(\omega)$ polarizability ellipsoids are not aligned and thus that $\mathbf{F}$ is non-vanishing in the strong-field limit. \textcolor{black}{For many molecules, it will suffice to choose $\omega^\prime$ somewhere in the near infrared and $\omega$ somewhere in the visible, thus avoiding vibrational absorption in the mid infrared and electronic absorption in the ultraviolet. In what follows, we consider $2\omega^\prime=\omega$ to facilitate experimental realization using a single laser source.}

\textcolor{black}{To calculate our chiral optical force $\mathbf{F}$, w}e take the molecule's rotational Hamiltonian to have the form
\begin{align}
H={}&V+U+H^{(0)}, \label{H}
\end{align}
where $V$ accounts for the molecule's interaction with the static field $\mathbf{E}_0$ and traveling-wave field $\tilde{\mathbf{E}}^\prime$, $U$ accounts for the molecule's interaction with the standing-wave field $\tilde{\mathbf{E}}$, and $H^{(0)}$ accounts for the \textcolor{black}{molecule's} rotational degrees of freedom in the absence of the fields. Ideally, we want $V$ to serve as a rotational potential energy and $U$ to serve as a translational potential energy, the field values being chosen such that $V\gg U$ \textcolor{black}{has the dominant effect on the molecule's orientation}. Neglecting nuclear spin \textcolor{black}{and modeling the molecule as an asymmetric rigid rotor}, we consider 
\begin{align}
V={}&-\mathcal{E}_z\mu_{0z}-\frac{1}{4}\mathcal{E}_x^{\prime 2}\alpha_{xx}(\omega^\prime), \nonumber \\
U={}&-\frac{1}{4}\mathcal{E}_y^2\alpha_{yy}(\omega)-\frac{1}{4}\mathcal{E}_x^2\alpha_{xx}(\omega) \nonumber \\
{}&-\frac{1}{2}\mathcal{E}_y\mathcal{E}_x\alpha_{yx}(\omega)\sin(2kZ_0), \nonumber \\
H^{(0)}={}&\frac{1}{\hbar^2}(CJ^2_c+BJ^2_b+AJ^2_a) \nonumber
\end{align}
with
\begin{align}
\mu_{0\alpha}={}&\ell_{\alpha\alpha^\prime}\mu_{0\alpha^\prime}, \nonumber \\
\alpha_{\beta\alpha}(\omega^\prime)={}&\ell_{\beta\beta^\prime}\ell_{\alpha\alpha^\prime}\alpha_{\beta^\prime\alpha^\prime}(\omega^\prime), \nonumber \\
\alpha_{\beta\alpha}(\omega)={}&\ell_{\beta\beta^\prime}\ell_{\alpha\alpha^\prime}\alpha_{\beta^\prime\alpha^\prime}(\omega), \nonumber
\end{align}
where \textcolor{black}{$\mathbf{R}_0$ is the molecule's position}; $C<B<A$ are the molecule's equilibrium rotational constants; $\mathbf{J}$ is the molecule's rotational angular momentum; $\ell$ is a direction cosine matrix relating the molecule's principal axes of inertia $c$, $b$, and $a$ to the laboratory axes $z$, $y$, and $x$; and the Einstein summation convention is to be understood with respect to indices $\dots,\beta^\prime,\alpha^\prime\in\{c,b,a\}$ and $\dots,\beta,\alpha\in\{z,y,x\}$ \cite{Barron04a, Bunker05a}. \textcolor{black}{The asymmetric rigid rotor is described in more detail in the appendices.} Let us assume that the molecule's rotational state has the form
\begin{align}
|\psi\rangle={}&\mathrm{e}^{-\mathrm{i}E_r t/\hbar}|r\rangle \nonumber
\end{align}
with
\begin{align}
H|r\rangle={}&E_r|r\rangle, \label{HrEr}
\end{align}
where $|r\rangle$ is a rotational energy eigenstate and $E_r$ is the associated energy eigenvalue. \textcolor{black}{It then follows that}
\begin{align}
\mathbf{F}={}&-\langle \psi|\partial_{Z_0}U|\psi\rangle\hat{\mathbf{z}} \nonumber \\
={}&k\mathcal{E}_y\mathcal{E}_x\textcolor{black}{\langle r|\alpha_{yx}(\omega)|r\rangle}\cos(2kZ_0)\hat{\mathbf{z}}. \label{GeneralF}
\end{align}
Note that $\mathbf{E}_0$ and $\tilde{\mathbf{E}}^\prime$ enter into Eq. \ref{GeneralF} implicitly via $|r\rangle$. 

Our analogy with \textit{Velella velella} \textcolor{black}{holds closest} in the strong-field regime ($|V|\gg|U|,H^{(0)}$), \textcolor{black}{where the molecule's rotation is heavily constrained by the static field $\mathbf{E}_0$ and traveling-wave field $\tilde{\mathbf{E}}^\prime$.} The lowest-lying rotational energy eigenstates $|r\rangle$ are then pendular states in which our chiral optical force tends towards
\begin{align}
\mathbf{F}\textcolor{black}{\rvert_{\Omega}={}}&k\mathcal{E}_y\mathcal{E}_x\alpha_{yx}(\omega)\rvert_\Omega\cos(2kZ_0)\hat{\mathbf{z}}, \label{StrongF}
\end{align}
where $\Omega$ denotes a molecular orientation for which the rotational potential energy $V$ is minimized. There are two such orientations, as $V$ has two-fold rotational symmetry\textcolor{black}{; an appropriate degree of orientation $D_r$ is defined in the appendices}. The sign of the orientated molecule's dipole moment component $\mu_{0z}\rvert_\Omega$ is dictated by the sign of the static field component $\mathcal{E}_z$ ($\mathrm{sgn}\,\mu_{0z}\rvert_\Omega=\mathrm{sgn}\,\mathcal{E}_z$) and the sign of the traveling-wave \textcolor{black}{frequency} polarizability component $\alpha_{yx}(\omega)\rvert_\Omega$ depends on both the sign of $\mu_{0z}\rvert_\Omega$ and the molecule's chirality. According to Eq. \ref{StrongF}, the direction of $\mathbf{F}\textcolor{black}{\rvert_\Omega}$ thus depends (implicitly) on the molecule's chirality via the sign of the product $[\mu_{0z}\alpha_{yx}(\omega)]\rvert_\Omega$ and on the field's chirality via the sign of the product $\mathcal{E}_z\mathcal{E}_y\mathcal{E}_x$. Note that with the molecule's orientation essentially fixed by $\mathbf{E}_0$ and $\tilde{\mathbf{E}}^\prime$ via $V$, the translational potential energy $U\rvert_\Omega$ is minimized when the polarization of the standing-wave field $\tilde{\mathbf{E}}$ is linear and quasi-parallel to the short axis of the projection of the $\alpha(\omega)$ polarizability ellipsoid in the $y$-$x$ plane. For $\alpha_{yx}(\omega)\rvert_\Omega\lessgtr0$, this occurs when $\sin(2kZ_0)=\mp1$; $\mathbf{F}$ pushes the molecule towards the nearest local minimum in $U\rvert_\Omega$. \textcolor{black}{The form of $\mathbf{F}$ in the weak-field regime and in the general case is considered in the appendices.}

\begin{figure}[h!]
\centering
\includegraphics[width=\linewidth]{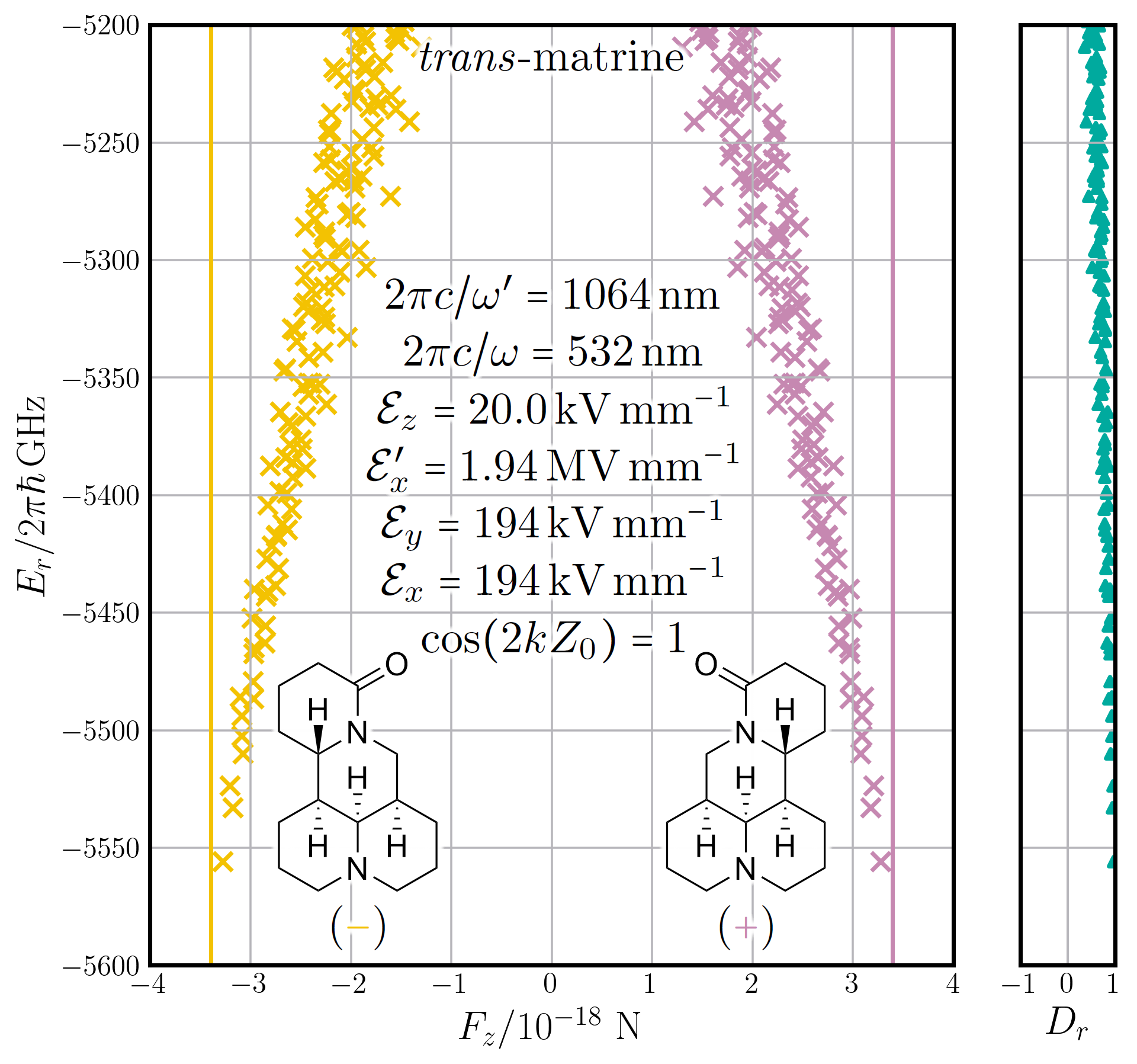}
\caption{\small Scatter plots of our chiral optical force $F_z$ \textcolor{black}{and degree of orientation $D_r$} versus rotational energy eigenvalue $E_r$ for some of the low-lying rotational states of ($\pm$)-\textit{trans}-matrine. \textcolor{black}{Each cross is a} quasi-degenerate ($5\%$ tolerance) \textcolor{black}{tunneling doublet}. Vertical lines indicate the strong-field limits $F_z\rvert_\Omega$. Note that the field has a fixed right-handed geometry ($\mathcal{E}_z\mathcal{E}_y\mathcal{E}_x>0$) with intensities of $I^\prime=\epsilon_0c\mathcal{E}_x^{\prime2}/2=5.00\times10^{11}\,\mathrm{W}\,\mathrm{cm}^{-2}$ and $I=\epsilon_0c(\mathcal{E}_y^2+\mathcal{E}_x^2)/2=1.00\times 10^{10}\,\mathrm{W}\,\mathrm{cm}^{-2}$.}
\label{Fig2}
\end{figure}

\textcolor{black}{Shown in Fig. \ref{Fig2} are results for ($\pm$)-\textit{trans}-matrine calculated as described in the appendices, where similar results can be seen for ($\pm$)-\textit{cis}-matrine. For the lowest energy rotational states, we approach the strong-field limits of $F_z\rvert_\Omega=\pm 3.32\times 10^{-18}\,\mathrm{N}$ given by Eq. \ref{StrongF} with the degree of orientation $D_r$ being close to $1$. The overall enantioselectivity is perfect in the range shown; $F_z$ has the same sign across rotational states for a given enantiomer and equal magnitudes but opposite signs for opposite enantiomers in a given rotational state. As the rotational energy $E_r$ increases, the strong-field criterion $|V|\gg H^{(0)}$ breaks down and $F_z$ deviates from $F_z\rvert_\Omega$ with a corresponding reduction of $D_r$. Samples of ($+$)-matrine can be extracted from traditional Chinese herbs and ($-$)-matrine can be readily synthesized, the \textit{trans}:\textit{cis} conformational ratio being around $1:0.30$ \cite{Juanes21a, Sun22a}. In contrast, Eq. \ref{Fh} gives $|\mathbf{F}^\prime|\le 6\times10^{-22}\,\mathrm{N}$ for matrine in a one-dimensional optical helicity lattice with commensurate parameters \cite{Cameron14b, Cameron14c}. Evidently then, our chiral optical force $\mathbf{F}$ can be over three orders of magnitude stronger than the corresponding optical helicity gradient force $\mathbf{F}^\prime$. We obtain similarly pleasing results for other typical chiral molecules. The special case of isotopically chiral molecules is considered in the appendices.}

Our \textcolor{black}{chiral optical force $\mathbf{F}$} can be regarded as part of the ongoing ``electric dipole revolution in chiral measurements’'\textcolor{black}{, joining photoelectron circular dichroism (PECD) \cite{Ritchie76a, Bowering01a, Kastner16a}, Coulomb explosion imaging \cite{Kitamura01a, Pitzer13a, Herwig13a}, enantiomer-specific microwave spectroscopy \cite{Hirota12a, Nafie13a, Patterson13a}, enantio-sensitive light bending \cite{Ayuso21a} and steering of free induction decay \cite{Khokhlova22a}, and other relatively new chiroptical phenomena based solely on electric dipole interactions \cite{Ayuso22a}. See \cite{Takemoto08a, Gershnabel18a, Ordonez23a} in particular for enantiomer-specific orientation schemes.}

\begin{figure}[h!]
\centering
\includegraphics[width=\linewidth]{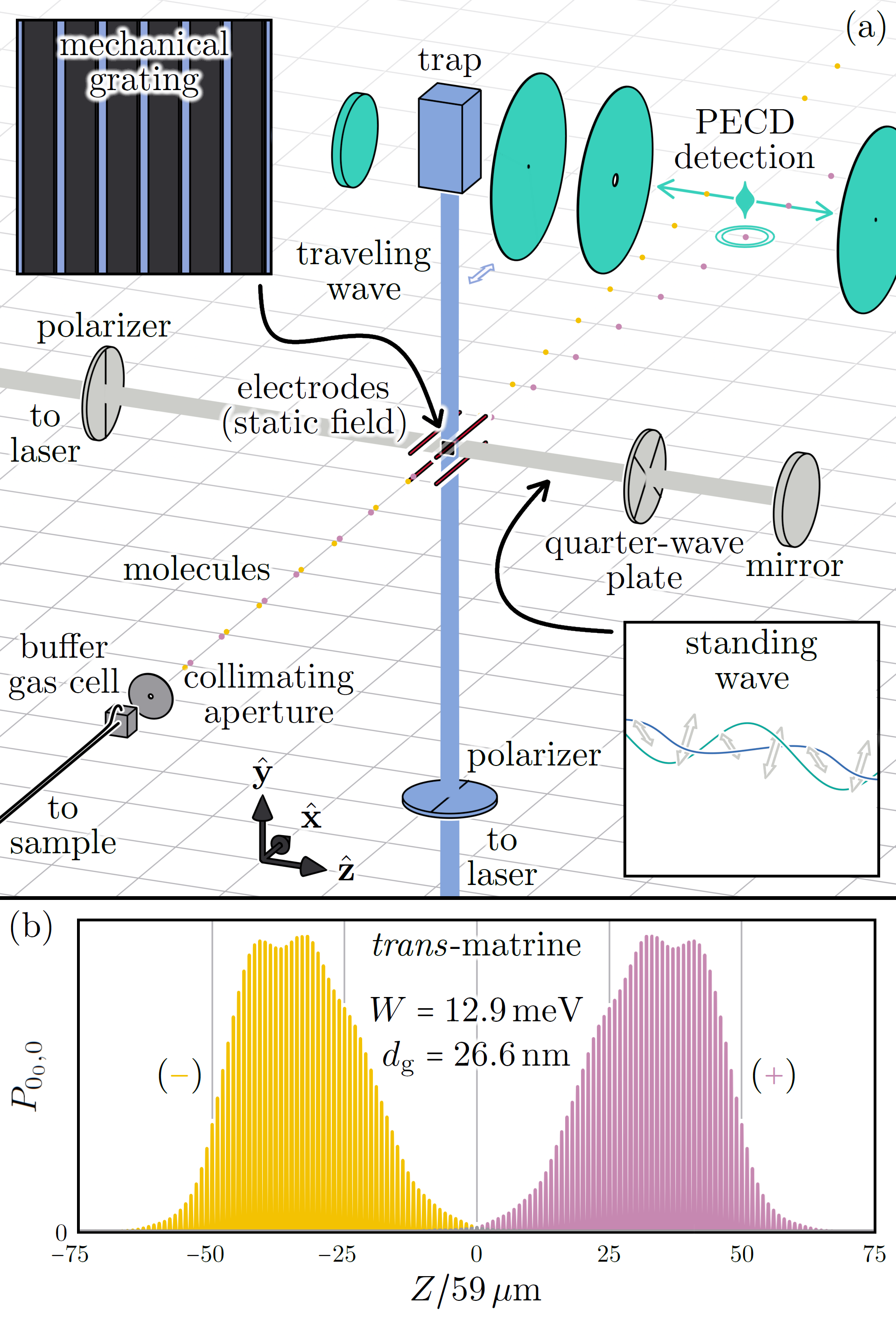}
\caption{\small \textcolor{black}{(a) A proposed experiment to separate the opposite enantiomers of a chiral molecule using our chiral optical force $\mathbf{F}$. (b) A predicted molecular deflection pattern. Note that $59\,\mu\mathrm{m}$ is the spacing between individual diffraction orders.}}
\label{Fig3}
\end{figure}

\textcolor{black}{We propose observing our chiral optical force $\mathbf{F}$ in a molecular beam deflection experiment as illustrated in Fig. \ref{Fig3}(a). A rigorous theoretical model of this experiment is presented in the appendices. The key features can be summarized as follows.}

\textcolor{black}{A cryogenically cooled helium buffer gas beam \cite{Hutzler12a, Samanta20a} of matrine enters an experimental chamber via a circular aperture of diameter $d_\mathrm{c}=1.00\,\mathrm{mm}$ at $x=x_\mathrm{c}=-1.10\,\mathrm{m}$. The ratio of molecules to helium atoms is $q\lesssim 1\times  10^{-2}$ and the chamber is evacuated to a residual pressure of $p\lesssim 10^{-8}\,\mathrm{mbar}$ at an ambient temperature of $T=300\,\mathrm{K}$. The molecules occupy their vibronic ground states and a Maxwell-Boltzmann distribution of their unperturbed rotational energy eigenstates at the cell temperature of $T_\mathrm{c}=4\,\mathrm{K}$. We consider continuous beam operation in the effusive to intermediate regime with a stagnation number density of helium in the cell of $n_{\mathrm{c},\mathrm{He}}\sim10^{14}$-$10^{15}\,\mathrm{cm}^{-3}$. Each molecule collides primarily with residual helium atoms in the experimental chamber at a mean rate of $\overline{\Gamma}_{\mathrm{m}-\mathrm{res}}\sim10^{-1}\,\mathrm{s}^{-1}$ \cite{Lubman82a, Patterson10a}. As the molecules are internally cold and essentially collision free, we describe their center-of-mass motion quantum mechanically in terms of matter waves \cite{Arndt99a, Brand21a}. A collimating aperture of diameter $1.00\,\mu\mathrm{m}$ at $x=x_\mathrm{a}=-1.00\,\mathrm{m}$ generates transverse matter-wave coherence downstream \cite{Brand21a}.}

\textcolor{black}{The static field $\mathbf{E}_0$ is generated by four cylindrical electrodes of length $l_\mathrm{e}=20\,\mathrm{cm}$ and diameter $d_\mathrm{e}=2\,\mathrm{mm}$ running parallel to the $x$ axis in a quadrupole geometry of width $w_\mathrm{e}=1\mathrm{mm}$ centered on $x=x_\mathrm{e}=0$. The electrodes are connected in pairs on either side of the $y$-$x$ plane to the terminals of a high-voltage dc power supply.}

\textcolor{black}{A thin mechanical grating at $x=x_\mathrm{g}=0$ mitigates the $\cos(2kZ)$ modulation of our chiral optical force $\mathbf{F}$ by only transmitting regions of the matter waves for which $\cos(2kZ)\ge0$ via $2N_\mathrm{g}+1=11$ slits of  height $h_\mathrm{g}=1.00\,\mu\mathrm{m}$ and width $26.6\,\mathrm{nm}=\pi/10k\le d_\mathrm{g}\le\pi/2k=133\,\mathrm{nm}$ arranged with a matching period of $\pi/k=266\,\mathrm{nm}$ along $z$ \cite{Arndt99a}. Together, the collimating aperture and mechanical grating provide longitudinal matter-wave coherence via grationally mediated spectral filtering (velocity selection) with fractional width $\Delta W_r/\overline{W}_r\sim 10^{-3}$ \cite{Brand21a}. The $\cos(2kZ)$ modulation could also be mitigated using spatially varying matter-wave depletion via optical absorption \cite{Walter16a}.}

\textcolor{black}{The travelling-wave field $\tilde{\mathbf{E}}^\prime$ and standing-wave field $\tilde{\mathbf{E}}$ are realized in pulsed form using a commercial Nd:YAG laser at the fundamental wavelength of $2\pi c/\omega^\prime=1064\,\mathrm{nm}$ and second-harmonic of $2\pi c/\omega=532\,\mathrm{nm}$. The laser pulses have Gaussian forms with spatial widths of $w_z^\prime=250\,\mu\mathrm{m}$ and $w_x^\prime=100\,\mu\mathrm{m}$ for the travelling wave and $w_y=250\,\mu\mathrm{m}$ and $w_x=100\,\mu\mathrm{m}$ for the standing wave with a common temporal width of $\tau_\mathrm{p}=10.0\,\mathrm{ns}$. They are focused at $x=x_\mathrm{p}\sim 1\,\mathrm{cm}$ at $t=t_\mathrm{p}=0$. The lin$\perp$lin configuration is formed by retroreflecting the second-harmonic laser pulse from a mirror combined with a quarter-wave plate.}

\textcolor{black}{Enantiosensitive detection is realized using PECD \cite{Ritchie76a, Bowering01a, Kastner16a} at $x=x_\mathrm{PECD}=1.00\,\mathrm{m}$ and $t=t_\mathrm{PECD}\sim10\,\mathrm{ms}$ with a pulse width of $w_\mathrm{PECD}=5.00\,\mathrm{\mu m}$ and duration of $\tau_\mathrm{PECD}=100\,\mathrm{fs}$.}

\textcolor{black}{Shown in Fig. \ref{Fig3}(b) are results for ($\pm$)-\textit{trans}-matrine calculated as described in the appendices, where similar results can be seen for ($\pm$)-\textit{cis}-matrine. A large separation ($\sim\mathrm{mm}$) of opposite enantiomers is immediately apparent.}

\textcolor{black}{This work could enable the first demonstration of a chiral optical force for individual chiral molecules and thus the first non-destructive enantioseparation using light.}

R.P.C. and D.M. gratefully acknowledge the support of the Royal Society (URF$\backslash$R1$\backslash$191243, RF$\backslash$ERE$\backslash$210170, RF$\backslash$ERE$\backslash$231130, and URF$\backslash$R$\backslash$241008). R.P.C. is a Royal Society University Research Fellow. N.V. and D.W. acknowledge the Deutsche Forschungsgemeinschaft for support (project 328961117) through the Collaborative Research Center ELCH (SFB 1319).

\onecolumngrid
\begin{appendix}

\section*{Asymmetric rigid rotor}
\label{Asymmetric rigid rotor}
The quantum-mechanical description of the asymmetric rigid rotor has been covered extensively elsewhere. We use the conventions below, which are slightly unusual in that we consider quantization along the $x$ axis rather than the $z$ axis.

We take the molecule's principal axes of inertia $c$, $b$, and $a$ to be related to the laboratory axes $z$, $y$, and $x$ via direction cosines given by
\begin{align}
\ell_{zc}={}&-\sin\chi\sin\phi\cos\theta+\cos\chi\cos\phi, \nonumber \\
\ell_{zb}={}&\cos\chi\sin\phi\cos\theta+\sin\chi\cos\phi, \nonumber \\
\ell_{za}={}&\sin\phi\sin\theta, \nonumber \\
\ell_{yc}={}&-\sin\chi\cos\phi\cos\theta-\cos\chi\sin\phi, \nonumber \\
\ell_{yb}={}&\cos\chi\cos\phi\cos\theta-\sin\chi\sin\phi, \nonumber \\
\ell_{ya}={}&\cos\phi\sin\theta, \nonumber \\
\ell_{xc}={}&\sin\chi\sin\theta, \nonumber \\
\ell_{xb}={}&-\cos\chi\sin\theta, \nonumber \\
\ell_{xa}={}&\cos\theta, \nonumber
\end{align}
where $0\le\chi<2\pi$, $0\le\phi<2\pi$, and $0\le\theta\le\pi$ are Euler angles. The molecule-fixed components of the molecule's rotational angular momentum $\mathbf{J}$ are
\begin{align}
J_c={}&-\mathrm{i}\hbar\left(-\sin\chi\cot\theta\frac{\partial}{\partial\chi}+\sin\chi\csc\theta\frac{\partial}{\partial\phi}+\cos\chi\frac{\partial}{\partial\theta}\right), \nonumber \\
J_b={}&-\mathrm{i}\hbar\left(\cos\chi\cot\theta\frac{\partial}{\partial\chi}-\cos\chi\csc\theta\frac{\partial}{\partial\phi}+\sin\chi\frac{\partial}{\partial\theta}\right), \nonumber \\
J_a={}&-\mathrm{i}\hbar\frac{\partial}{\partial\chi}. \nonumber
\end{align}
The laboratory-fixed components of $\mathbf{J}$ are
\begin{align}
J_z={}&-\mathrm{i}\hbar\left(\sin\phi\csc\theta\frac{\partial}{\partial\chi}-\sin\phi\cot\theta\frac{\partial}{\partial\phi}+\cos\phi\frac{\partial}{\partial\theta}\right), \nonumber \\
J_y={}&-\mathrm{i}\hbar\left(\cos\phi\csc\theta\frac{\partial}{\partial\chi}-\cos\phi\cot\theta\frac{\partial}{\partial\phi}-\sin\phi\frac{\partial}{\partial\theta}\right), \nonumber \\
J_x={}&-\mathrm{i}\hbar\frac{\partial}{\partial\phi}. \nonumber
\end{align}
We take the unperturbed rotational energy eigenstates $|J_{\tau},m\rangle^{(0)}$ and associated energy eigenvalues $E_{J_\tau}^{(0)}$ to satisfy
\begin{align}
H^{(0)}|J_\tau,m\rangle^{(0)}={}&E_{J_\tau}^{(0)}|J_\tau,m\rangle^{(0)}, \nonumber \\
(J_c^2+J_b^2+J_a^2)|J_\tau,m\rangle^{(0)}={}&J(J+1)\hbar^2|J_\tau,m\rangle^{(0)}, \nonumber \\
J_x|J_\tau,m\rangle^{(0)}={}&m\hbar|J_\tau,m\rangle^{(0)}, \nonumber
\end{align}
where $J\in\{0,1,\dots\}$ is the rotational angular momentum quantum number, $\tau\in\{-J,\dots,J\}$ is a label that increases with increasing energy, and $m\in\{-J,\dots,J\}$ is the laboratory-fixed rotational angular momentum projection quantum number.

To help us identify the unperturbed rotational energy eigenstates $|J_{\tau},m\rangle^{(0)}$ and associated energy eigenvalues $E_{J_\tau}^{(0)}$ explicitly, we work in a basis of unperturbed symmetric rigid rotor states $|J,\kappa,m\rangle^{(0)}$ satisfying
\begin{align}
(J_c^2+J_b^2+J_a^2)|J,\kappa,m\rangle^{(0)}={}&J(J+1)\hbar^2|J,\kappa,m\rangle^{(0)}, \nonumber \\
J_a|J,\kappa,m\rangle^{(0)}={}&\kappa\hbar|J,\kappa,m\rangle^{(0)}, \nonumber \\
J_x|J,\kappa,m\rangle^{(0)}={}&m\hbar|J,\kappa,m\rangle^{(0)}, \nonumber
\end{align}
where $\kappa\in\{-J,\dots,J\}$ is the molecule-fixed rotational angular momentum projection quantum number, the corresponding wavefunctions $\langle\chi,\phi,\theta|J,\kappa,m\rangle^{(0)}$ being given by
\begin{align}
\langle\chi,\phi,\theta|J,\kappa,m\rangle^{(0)}={}&\sqrt{\frac{(2J+1)(J+\kappa)!(J-\kappa)!(J+m)!(J-m)!}{8\pi^2}}\mathrm{e}^{\mathrm{i}\kappa\chi}\mathrm{e}^{\mathrm{i}m\phi} \nonumber \\
{}&\times\sum_{\sigma=\mathrm{max}(0,\kappa-m)}^{\mathrm{min}(J+\kappa,J-m)}(-1)^{-\kappa+m+\sigma}\frac{[\cos(\theta/2)]^{2J+\kappa-m-2\sigma}[\sin(\theta/2)]^{-\kappa+m+2\sigma}}{(J+\kappa-\sigma)!(J-m-\sigma)!(-\kappa+m+\sigma)!\sigma!}. \nonumber
\end{align}
The $|J,\kappa,m\rangle^{(0)}$ render the unperturbed rotational Hamiltonian $H^{(0)}$ block diagonal in the quantum number $J$, as
\begin{align}
{}^{(0)}\langle J^\prime,\kappa^\prime,m^\prime|H^{(0)}|J,\kappa,m\rangle^{(0)}={}&\frac{1}{2}J(J+1)(C+B)\delta_{J^\prime J}\delta_{\kappa^\prime\kappa}\delta_{m^\prime m}-\kappa^2\left[\frac{1}{2}(C+B)-A\right]\delta_{J^\prime J}\delta_{\kappa^\prime\kappa}\delta_{m^\prime m} \nonumber \\
{}&-\frac{1}{4}\sqrt{[J(J+1)-\kappa(\kappa-1)][J(J+1)-(\kappa-1)(\kappa-2)]}(C-B)\delta_{J^\prime J}\delta_{\kappa^\prime\kappa-2}\delta_{m^\prime m} \nonumber \\
{}&-\frac{1}{4}\sqrt{[J(J+1)-\kappa(\kappa+1)][J(J+1)-(\kappa+1)(\kappa+2)]}(C-B)\delta_{J^\prime J}\delta_{\kappa^\prime\kappa+2}\delta_{m^\prime m}. \nonumber
\end{align}
The diagonalization of $H^{(0)}$ can completed analytically for $J\in\{0,\dots,5\}$, giving
\begin{align}
|0_0,0\rangle^{(0)}={}&|0,0,0\rangle^{(0)}, \nonumber \\
|1_{-1},m\rangle^{(0)}={}&|1,0,m\rangle^{(0)}, \nonumber \\
|1_0,m\rangle^{(0)}={}&\frac{1}{\sqrt{2}}(|1,1,m\rangle^{(0)}-|1,-1,m\rangle^{(0)}), \nonumber \\
|1_1,m\rangle^{(0)}={}&\frac{1}{\sqrt{2}}(|1,1,m\rangle^{(0)}+|1,-1,m\rangle^{(0)})  \nonumber
\end{align}
with
\begin{align}
E_{0_0}^{(0)}={}&0, \nonumber \\
E_{1_{-1}}^{(0)}={}&C+B, \nonumber \\
E_{1_0}^{(0)}={}&C+A, \nonumber \\
E_{1_1}^{(0)}={}&B+A, \nonumber
\end{align}
for example. The diagonalization of $H^{(0)}$ must be completed numerically for $J\in\{6,\dots\}$.

In the symmetric rotor basis, we have
\begin{align}
{}^{(0)}\langle J^\prime,\kappa^\prime,m^\prime|\ell_{zc}|J,\kappa,m\rangle^{(0)}={}&H_{J^\prime J}G_{J^\prime\kappa^\prime J\kappa}F_{J^\prime m^\prime Jm}(-\mathrm{i}\delta_{\kappa^\prime\kappa-1}+\mathrm{i}\delta_{\kappa^\prime\kappa+1})(\mathrm{i}\delta_{m^\prime m-1}-\mathrm{i}\delta_{m^\prime m+1}), \nonumber \\
{}^{(0)}\langle J^\prime,\kappa^\prime,m^\prime|\ell_{zb}|J,\kappa,m\rangle^{(0)}={}&H_{J^\prime J}G_{J^\prime\kappa^\prime J\kappa}F_{J^\prime m^\prime Jm}(\delta_{\kappa^\prime\kappa-1}+\delta_{\kappa^\prime\kappa+1})(\mathrm{i}\delta_{m^\prime m-1}-\mathrm{i}\delta_{m^\prime m+1}), \nonumber \\
{}^{(0)}\langle J^\prime,\kappa^\prime,m^\prime|\ell_{za}|J,\kappa,m\rangle^{(0)}={}&H_{J^\prime J}G_{J^\prime\kappa^\prime J\kappa}F_{J^\prime m^\prime Jm}\delta_{\kappa^\prime\kappa}(\mathrm{i}\delta_{m^\prime m-1}-\mathrm{i}\delta_{m^\prime m+1}), \nonumber \\
{}^{(0)}\langle J^\prime,\kappa^\prime,m^\prime|\ell_{yc}|J,\kappa,m\rangle^{(0)}={}&H_{J^\prime J}G_{J^\prime\kappa^\prime J\kappa}F_{J^\prime m^\prime Jm}(-\mathrm{i}\delta_{\kappa^\prime\kappa-1}+\mathrm{i}\delta_{\kappa^\prime\kappa+1})(\delta_{m^\prime m-1}+\delta_{m^\prime m+1}), \nonumber \\
{}^{(0)}\langle J^\prime,\kappa^\prime,m^\prime|\ell_{yb}|J,\kappa,m\rangle^{(0)}={}&H_{J^\prime J}G_{J^\prime\kappa^\prime J\kappa}F_{J^\prime m^\prime Jm}(\delta_{\kappa^\prime\kappa-1}+\delta_{\kappa^\prime\kappa+1})(\delta_{m^\prime m-1}+\delta_{m^\prime m+1}), \nonumber \\
{}^{(0)}\langle J^\prime,\kappa^\prime,m^\prime|\ell_{ya}|J,\kappa,m\rangle^{(0)}={}&H_{J^\prime J}G_{J^\prime\kappa^\prime J\kappa}F_{J^\prime m^\prime Jm}\delta_{\kappa^\prime\kappa}(\delta_{m^\prime m-1}+\delta_{m^\prime m+1}), \nonumber \\
{}^{(0)}\langle J^\prime,\kappa^\prime,m^\prime|\ell_{xc}|J,\kappa,m\rangle^{(0)}={}&H_{J^\prime J}G_{J^\prime\kappa^\prime J\kappa}F_{J^\prime m^\prime Jm}(-\mathrm{i}\delta_{\kappa^\prime\kappa-1}+\mathrm{i}\delta_{\kappa^\prime\kappa+1})\delta_{m^\prime m}, \nonumber \\
{}^{(0)}\langle J^\prime,\kappa^\prime,m^\prime|\ell_{xb}|J,\kappa,m\rangle^{(0)}={}&H_{J^\prime J}G_{J^\prime\kappa^\prime J\kappa}F_{J^\prime m^\prime Jm}(\delta_{\kappa^\prime\kappa-1}+\delta_{\kappa^\prime\kappa+1})\delta_{m^\prime m}, \nonumber \\
{}^{(0)}\langle J^\prime,\kappa^\prime,m^\prime|\ell_{xa}|J,\kappa,m\rangle^{(0)}={}&H_{J^\prime J}G_{J^\prime\kappa^\prime J\kappa}F_{J^\prime m^\prime Jm}\delta_{\kappa^\prime\kappa}\delta_{m^\prime m}, \nonumber
\end{align}
\textcolor{black}{where}
\begin{align}
H_{J^\prime J}={}&\frac{1}{4(J+\delta_{0J})\sqrt{(2J-1)(2J+1)}}\delta_{J^\prime J-1}+\frac{1}{4(J+\delta_{0J})(J+1)}\delta_{J^\prime J}+\frac{1}{4(J+1)\sqrt{4(J+1)^2-1}}\delta_{J^\prime J+1}, \nonumber \\
G_{J^\prime\kappa^\prime J\kappa}={}&-\sqrt{(J+\kappa-1)(J+\kappa)}\delta_{J^\prime J-1}\delta_{\kappa^\prime\kappa-1} +\sqrt{(J-\kappa+1)(J+\kappa)}\delta_{J^\prime J}\delta_{\kappa^\prime\kappa-1}+\sqrt{(J-\kappa+2)(J-\kappa+1)}\delta_{J^\prime J+1}\delta_{\kappa^\prime\kappa-1} \nonumber \\
{}&+2\sqrt{J^2-\kappa^2}\delta_{J^\prime J-1}\delta_{\kappa^\prime\kappa}+2\kappa\delta_{J^\prime J}\delta_{\kappa^\prime\kappa}+2\sqrt{(J+1)^2-\kappa^2}\delta_{J^\prime J+1}\delta_{\kappa^\prime\kappa}+\sqrt{(J-\kappa-1)(J-\kappa)}\delta_{J^\prime J-1}\delta_{\kappa^\prime\kappa+1} \nonumber \\
{}&+\sqrt{(J+\kappa+1)(J-\kappa)}\delta_{J^\prime J}\delta_{\kappa^\prime\kappa+1} -\sqrt{(J+\kappa+2)(J+\kappa+1)}\delta_{J^\prime J+1}\delta_{\kappa^\prime\kappa+1}, \nonumber \\
F_{J^\prime m^\prime Jm}={}&-\sqrt{(J+m-1)(J+m)}\delta_{J^\prime J-1}\delta_{m^\prime m-1}+\sqrt{(J-m+1)(J+m)}\delta_{J^\prime J}\delta_{m^\prime m-1} \nonumber \\
{}&+\sqrt{(J-m+2)(J-m+1)}\delta_{J^\prime J+1}\delta_{m^\prime m-1}+2\sqrt{J^2-m^2}\delta_{J^\prime J-1}\delta_{m^\prime m}+2m\delta_{J^\prime J}\delta_{m^\prime m} \nonumber \\
{}&+2\sqrt{(J+1)^2-m^2}\delta_{J^\prime J+1}\delta_{m^\prime m}+\sqrt{(J-m-1)(J-m)}\delta_{J^\prime J-1}\delta_{m^\prime m+1}+\sqrt{(J+m+1)(J-m)}\delta_{J^\prime J}\delta_{m^\prime m+1} \nonumber \\
{}&-\sqrt{(J+m+2)(J+m+1)}\delta_{J^\prime J+1}\delta_{m^\prime m+1} \nonumber
\end{align}
are useful shorthands.

\section*{Weak-field regime}
\label{Weak-field regime}
\textcolor{black}{In the weak-field regime ($H^{(0)}\gg|V|\gg|U|$), the molecule's rotation is only slightly affected by the static field $\mathbf{E}_0$ and traveling-wave field $\tilde{\mathbf{E}}^\prime$.} Basic perturbation theory valid to first order in the rotational potential energy $V$ with the translational potential energy $U$ neglected gives
\begin{align}
|J_\tau,m\rangle={}&|J_\tau,m\rangle^{(0)}+|J_\tau,m\rangle^{(1)}, \nonumber
\end{align}
\textcolor{black}{where}
\begin{align}
|J_\tau,m\rangle^{(1)}={}&\sum_{\substack{\{J^\prime_{\tau^\prime},m^\prime\}\\(J^\prime_{\tau^\prime}\ne J_\tau)}}\frac{{}^{(0)}\langle J^\prime_{\tau^\prime},m^\prime|V|J_\tau,m\rangle^{(0)}}{E_{J_\tau}^{(0)}-E_{J^\prime_{\tau^\prime}}^{(0)}}|J^\prime_{\tau^\prime},m^\prime\rangle^{(0)} \nonumber
\end{align}
is the first-order correction to the unperturbed rotational energy eigenstate $|J_\tau,m\rangle^{(0)}$. Note that \textcolor{black}{the matrix} $^{(0)}\langle J_\tau,m^\prime|V|J_\tau,m\rangle^{(0)}\propto\delta_{m^\prime m}$ is diagonal in the quantum number $m$, as is required for the self-consistent application of the perturbation theory; the asymmetric rigid rotor has no linear dc Stark shift and the $|J_\tau,m\rangle^{(0)}$ are quantized along $x$, matching the polarization axis of $\tilde{\mathbf{E}}^\prime$. To first order in $V$, the expectation value required to evaluate our chiral optical force $\mathbf{F}$ is
\begin{align}
\langle {}&J_\tau,m|\ell_{y\beta^\prime}\ell_{x\alpha^\prime}|J_\tau,m\rangle\alpha_{\beta^\prime\alpha^\prime}(\omega)={}^{(0)}\langle J_\tau,m|\ell_{y\beta^\prime}\ell_{x\alpha^\prime}|J_\tau,m\rangle^{(0)}\alpha_{\beta^\prime\alpha^\prime}(\omega) \nonumber \\
{}&+\sum_{\substack{\{J^\prime_{\tau^\prime},m^\prime\}\\(J^\prime_{\tau^\prime}\ne J_\tau)}}\frac{2\mathcal{E}_z}{E_{J^\prime_{\tau^\prime}}^{(0)}-E_{J_\tau}^{(0)}}\Re({}^{(0)}\langle J_\tau,m|\ell_{z\gamma^\prime}|J^\prime_{\tau^\prime},m^\prime\rangle^{(0)}{}^{(0)}\langle J^\prime_{\tau^\prime},m^\prime|\ell_{y\beta^\prime}\ell_{x\alpha^\prime}|J_\tau,m\rangle^{(0)})\mu_{0\gamma^\prime}\alpha_{\beta^\prime\alpha^\prime}(\omega) \nonumber \\
{}&+\sum_{\substack{\{J^\prime_{\tau^\prime},m^\prime\}\\(J^\prime_{\tau^\prime}\ne J_\tau)}}\frac{\mathcal{E}_x^{\prime 2}}{2(E_{J^\prime_{\tau^\prime}}^{(0)}-E_{J_\tau}^{(0)})}\Re({}^{(0)}\langle J_\tau,m|\ell_{x\delta^\prime}\ell_{x\gamma^\prime}|J^\prime_{\tau^\prime},m^\prime\rangle^{(0)}{}^{(0)}\langle J^\prime_{\tau^\prime},m^\prime|\ell_{y\beta^\prime}\ell_{x\alpha^\prime}|J_\tau,m\rangle^{(0)})\alpha_{\delta^\prime\gamma^\prime}(\omega^\prime)\alpha_{\beta^\prime\alpha^\prime}(\omega). \label{FullPerturbative}
\end{align}
Explicit calculation reveals that the first and final terms in Eq. \ref{FullPerturbative} vanish. As $\mu_{0c}\alpha_{ba}(\omega)=\mu_{0c}\alpha_{ab}(\omega)$, $\mu_{0b}\alpha_{ac}(\omega)=\mu_{0b}\alpha_{ca}(\omega)$, and $\mu_{0a}\alpha_{cb}(\omega)=\mu_{0a}\alpha_{bc}(\omega)$ are the only components of the product $\mu_{0\gamma^\prime}\alpha_{\beta^\prime\alpha^\prime}(\omega)$ that are uniquely signed and thus directly observable for the asymmetric rigid rotor, \textcolor{black}{we have, to first order in $V$,}
\begin{align}
\mathbf{F}={}&k\mathcal{E}_z\mathcal{E}_y\mathcal{E}_x[\mathcal{C}_r\mu_{0c}\alpha_{ba}(\omega)+\mathcal{B}_r\mu_{0b}\alpha_{ac}(\omega)+\mathcal{A}_r\mu_{0a}\alpha_{cb}(\omega)]\cos(2kZ_0)\hat{\mathbf{z}}, \label{PerturbativeF}
\end{align}
\textcolor{black}{where}
\begin{align}
\mathcal{C}_r={}&\sum_{\substack{\{J^\prime_{\tau^\prime},m^\prime\}\\(J^\prime_{\tau^\prime}\ne J_\tau)}}\frac{2}{E_{J^\prime_{\tau^\prime}}^{(0)}-E_{J_\tau}^{(0)}}\Re({}^{(0)}\langle J_\tau,m|\ell_{zc}|J^\prime_{\tau^\prime},m^\prime\rangle^{(0)}{}^{(0)}\langle J^\prime_{\tau^\prime},m^\prime|(\ell_{yb}\ell_{xa}+\ell_{ya}\ell_{xb})|J_\tau,m\rangle^{(0)}), \nonumber \\
\mathcal{B}_r={}&\sum_{\substack{\{J^\prime_{\tau^\prime},m^\prime\}\\(J^\prime_{\tau^\prime}\ne J_\tau)}}\frac{2}{E_{J^\prime_{\tau^\prime}}^{(0)}-E_{J_\tau}^{(0)}}\Re({}^{(0)}\langle J_\tau,m|\ell_{zb}|J^\prime_{\tau^\prime},m^\prime\rangle^{(0)}{}^{(0)}\langle J^\prime_{\tau^\prime},m^\prime|(\ell_{ya}\ell_{xc}+\ell_{yc}\ell_{xa})|J_\tau,m\rangle^{(0)}), \nonumber \\
\mathcal{A}_r={}&\sum_{\substack{\{J^\prime_{\tau^\prime},m^\prime\}\\(J^\prime_{\tau^\prime}\ne J_\tau)}}\frac{2}{E_{J^\prime_{\tau^\prime}}^{(0)}-E_{J_\tau}^{(0)}}\Re({}^{(0)}\langle J_\tau,m|\ell_{za}|J^\prime_{\tau^\prime},m^\prime\rangle^{(0)}{}^{(0)}\langle J^\prime_{\tau^\prime},m^\prime|(\ell_{yc}\ell_{xb}+\ell_{yb}\ell_{xc})|J_\tau,m\rangle^{(0)}) \nonumber
\end{align}
are chirally insensitive coefficients, the values of which depend on the molecule's rotational state \cite{Cameron23a}. According to Eq. \ref{PerturbativeF}, the direction of $\mathbf{F}$ depends on the molecule's chirality via the products $\mu_{0c}\alpha_{ba}(\omega)$, $\mu_{0b}\alpha_{ac}(\omega)$, and $\mu_{0a}\alpha_{cb}(\omega)$ and on the field's chirality via the product $\mathcal{E}_z\mathcal{E}_y\mathcal{E}_x$.

\section*{General case}
\label{General case}
\textcolor{black}{Between the extremes of the strong-field limit given by Eq. \ref{StrongF} and the weak-field limit given by Eq. \ref{PerturbativeF}, our chiral optical force $\mathbf{F}$ has to be calculated using Eq. \ref{GeneralF} with the rotational energy eigenstate $|r\rangle$ obtained numerically. In general, we find that the associated energy eigenvalue and required expectation value have the forms
\begin{align}
E_r={}&-\Delta_r-\frac{1}{2}\mathcal{E}_y\mathcal{E}_x\chi_r\sin(2kZ_0)-\frac{1}{2}\mathcal{E}_y\mathcal{E}_x\sum_{n=1}^\infty\frac{1}{(n+1)}\Upsilon_r^{(n)}\sin^{n+1}(2kZ_0) \label{Er} \\
\langle r|\alpha_{yx}(\omega)|r\rangle={}&\chi_r+\sum_{n=1}^\infty\Upsilon_r^{(n)}\sin^{n}(2kZ_0) \nonumber
\end{align}
giving
\begin{align}
\mathbf{F}={}&k\mathcal{E}_y\mathcal{E}_x\left[\chi_r+\sum_{n=1}^\infty\Upsilon_r^{(n)}\sin^{n}(2kZ_0)\right]\cos(2kZ_0)\hat{\mathbf{z}}, \nonumber
\end{align}
where $\Delta_r$ is an energy offset, $\chi_r$ is the leading effective polarizability, and the $\Upsilon_r^{(n)}$ embody corrections due to the unwanted effect of the translational potential energy $U$ on the molecule's orientation. Symmetry arguments dictate that  $\Delta_r$ and the odd $\Upsilon_r^{(n)}$ are chirally insensitive whereas $\chi_r$ and the even $\Upsilon_r^{(n)}$ have opposite signs for opposite enantiomers. The values of $\Delta_r$, $\chi_r$, and the $\Upsilon_r^{(n)}$ depend on the molecule's rotational state. Note that $\mathbf{F}=-\partial_{Z_0}E_r\hat{\mathbf{z}}$ and that the $\Upsilon_r^{(n)}$ contribute nothing to $\mathbf{F}$ when $\sin(2kZ_0)=0$.}

\section*{Degree of orientation}
\label{Degree of orientation}
\textcolor{black}{Let $\Omega_1$ and $\Omega_2$ denote the molecular orientations that minimize the rotational potential energy $V$ ($\Omega\in\{\Omega_1,\Omega_2\}$); $\Omega_1$ and $\Omega_2$ are antiorientations of each other related by a $\pi$ rotation about the $z$ axis. Extrapolating from \cite{Makhija12a}, we quantify the degree of molecular orientation in rotational energy eigenstate $|r\rangle$ by calculating
\begin{align}
D_r={}&-\langle r|\cos\delta_{\Omega_1}\cos\delta_{\Omega_2}|r\rangle \nonumber \\
={}&-\frac{1}{4}\langle r|\ell_{\alpha\alpha^\prime}\ell_{\beta\beta^\prime}|r\rangle\ell_{\alpha\alpha^\prime}(\Omega_1)\ell_{\beta\beta^\prime}(\Omega_2)+\frac{1}{4}\langle r|\ell_{\alpha\alpha^\prime}|r\rangle\ell_{\alpha\alpha^\prime}(\Omega_1)+\frac{1}{4}\langle r|\ell_{\alpha\alpha^\prime}|r\rangle\ell_{\alpha\alpha^\prime}(\Omega_2)-\frac{1}{4}, \label{Dr}
\end{align}
where $0\le\delta_{\Omega_1}\le\pi$ ($0\le\delta_{\Omega_2}\le\pi$) is the single-axis angle through which the molecule must be rotated to reach the target orientation $\Omega_1$ ($\Omega_2$) from an initial orientation with Euler angles $\chi$, $\phi$, and $\theta$. The degree of orientation $D_r$ equals $1$ for the optimal case where $|r\rangle$ is entirely localised on $\Omega_1$ and/or $\Omega_2$ and $-1$ when this $|r\rangle$ is rotated by $\pi$ about any axis perpendicular to the $z$ axis.}

\section*{Scatter plots}
\label{Scatter plots}
\textcolor{black}{To produce scatter plots of our chiral optical force $F_z$ and degree of orientation $D_r$ versus rotational energy eigenvalue $E_r$, we first calculated the molecular properties quoted below with the aid of Gaussian 09 \cite{Gaussian09a}. For matrine, we used the optimized geometries quoted in \cite{Juanes21a}, these having been calculated at the DFT B3LYP-D3(BJ)/def2-TZVP level of theory. For eucalyptol and isotopically chiral eucalyptol, we calculated the optimized geometries at the DFT B3LYP/6-311++G(d,p) level of theory. For each optimized geometry, we calculated the rotational constants $C$, $B$, and $A$ and then the dipole moment $\pmb{\mu}_0$\textcolor{black}{,} traveling-wave \textcolor{black}{frequency} polarizability $\alpha(\omega^\prime)$, and standing-wave \textcolor{black}{frequency} polarizability $\alpha(\omega)$ at the DFT B3LYP/aug-cc-pVDZ level of theory. For each set of molecular properties, we calculated the rotational Hamiltonian $H$ in the symmetric rotor basis truncated to $J\in\{0,\dots,16\}$ \textcolor{black}{and performed a diagonalization to obtain} the rotational energy eigenstates $|r\rangle$ and associated energy eigenvalues $E_r$. For each $|r\rangle$ and $E_r$ pair, we calculated $\mathbf{F}$ according to Eq. \ref{GeneralF} and $D_r$ according to Eq. \ref{Dr}.}

\textcolor{black}{For matrine with isotopic constitution ${}^{12}\mathrm{C}_{15}{}^1\mathrm{H}_{24}{}^{14}\mathrm{N}_2{}^{16}\mathrm{O}$, we obtained
\begin{align}
\frac{1}{2\pi\hbar}\left[\begin{matrix}
C & B & A
\end{matrix}\right]={}&\left[\begin{matrix}
0.26349 & 0.37394 & 0.67093
\end{matrix}\right]\,\mathrm{GHz}, \nonumber \\
\left[\begin{matrix}
\mu_{0c} & \mu_{0b} & \mu_{0a}
\end{matrix}\right]={}&\left[\begin{matrix}
\pm0.24846 & \mp12.48837 & \pm8.27507
\end{matrix}\right]\times10^{-30}\,\mathrm{C}\,\mathrm{m}, \nonumber \\
\left[\begin{matrix}
\alpha_{cc} & \alpha_{cb} & \alpha_{ca} \\
\alpha_{bc} & \alpha_{bb} & \alpha_{ba} \\
\alpha_{ac} & \alpha_{ab} & \alpha_{aa}
\end{matrix}\right](\omega^\prime)={}&\left[\begin{matrix}
2479.36 & 2.16 & -7.43 \\
2.16 & 3240.47 & -21.56 \\
-7.43 & -21.56 & 3608.42
\end{matrix}\right]\times10^{-42}\,\mathrm{C}^2\,\mathrm{m}^2\,\mathrm{J}^{-1}, \nonumber \\
\left[\begin{matrix}
\alpha_{cc} & \alpha_{cb} & \alpha_{ca} \\
\alpha_{bc} & \alpha_{bb} & \alpha_{ba} \\
\alpha_{ac} & \alpha_{ab} & \alpha_{aa}
\end{matrix}\right](\omega)={}&\left[\begin{matrix}
2538.92 & 0.55 & 0.46 \\
0.55 & 3316.27 & -31.48 \\
0.46 & -31.48 & 3722.23
\end{matrix}\right]\times10^{-42}\,\mathrm{C}^2\,\mathrm{m}^2\,\mathrm{J}^{-1}, \nonumber
\end{align}
for ($\pm$)-\textit{trans}-matrine and
\begin{align}
\frac{1}{2\pi\hbar}\left[\begin{matrix}
C & B & A
\end{matrix}\right]={}&\left[\begin{matrix}
0.27154 & 0.39178 & 0.67555
\end{matrix}\right]\,\mathrm{GHz}, \nonumber \\
\left[\begin{matrix}
\mu_{0c} & \mu_{0b} & \mu_{0a}
\end{matrix}\right]={}&\left[\begin{matrix}
\mp0.52888 & \mp12.28293 & \pm7.81973
\end{matrix}\right]\times10^{-30}\,\mathrm{C}\,\mathrm{m}, \nonumber \\
\left[\begin{matrix}
\alpha_{cc} & \alpha_{cb} & \alpha_{ca} \\
\alpha_{bc} & \alpha_{bb} & \alpha_{ba} \\
\alpha_{ac} & \alpha_{ab} & \alpha_{aa}
\end{matrix}\right](\omega^\prime)={}&\left[\begin{matrix}
2468.61 & 17.99 & -27.18 \\
17.99 & 3227.00 & -7.25 \\
-27.18 & -7.25 & 3561.47
\end{matrix}\right]\times10^{-42}\,\mathrm{C}^2\,\mathrm{m}^2\,\mathrm{J}^{-1}, \nonumber \\
\left[\begin{matrix}
\alpha_{cc} & \alpha_{cb} & \alpha_{ca} \\
\alpha_{bc} & \alpha_{bb} & \alpha_{ba} \\
\alpha_{ac} & \alpha_{ab} & \alpha_{aa}
\end{matrix}\right](\omega)={}&\left[\begin{matrix}
2529.08 & 17.10 & -19.83 \\
17.10 & 3302.34 & -15.33 \\
-19.83 & -15.33 & 3669.25
\end{matrix}\right]\times10^{-42}\,\mathrm{C}^2\,\mathrm{m}^2\,\mathrm{J}^{-1}, \nonumber
\end{align}
for ($\pm$)-\textit{cis}-matrine.}

\begin{figure}[h!]
\centering
\includegraphics[width=0.5\linewidth]{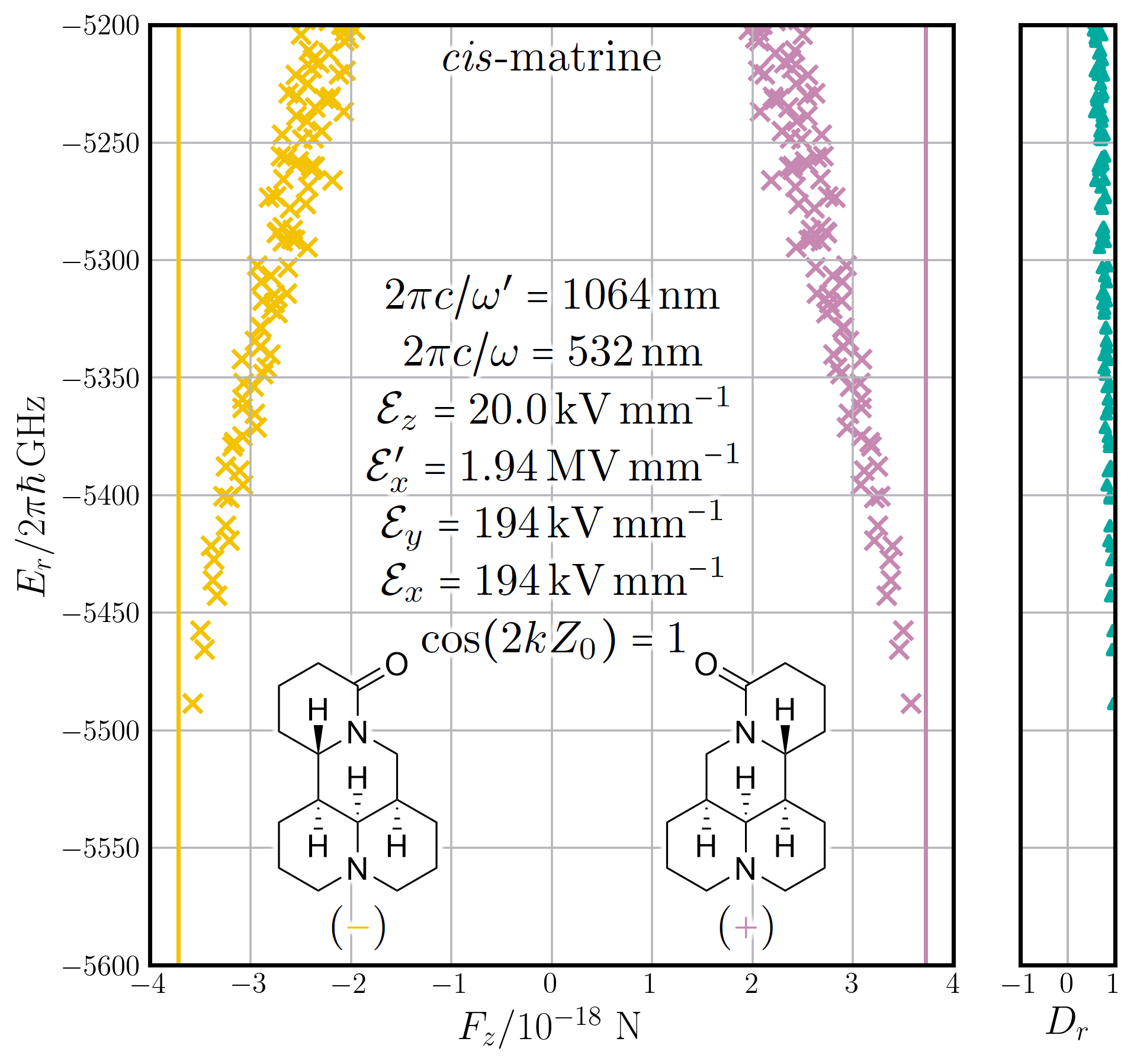}
\caption{\small As in Fig. \ref{Fig2} but for the \textit{cis} conformation.}
\label{Fig4}
\end{figure}

\textcolor{black}{Shown in Fig. \ref{Fig4} are results for ($\pm$)-\textit{cis}-matrine. The qualitative similarity to Fig. \ref{Fig2} illustrates the fact that our chiral optical force $\mathbf{F}$ is robust against small changes in molecular geometry. As a further check on the reliability of our predictions, we compared the strong-field limits of $F_z\rvert_\Omega=\pm 3.32\times 10^{-18}\,\mathrm{N}$ for ($\pm$)-\textit{trans}-matrine and $F_z\rvert_\Omega=\pm 3.73\times 10^{-18}\,\mathrm{N}$ for ($\pm$)-\textit{cis}-matrine with the values obtained using other functionals and basis sets, obtaining $F_z\rvert_\Omega=\pm 2.99\times 10^{-18}\,\mathrm{N}$ and $F_z\rvert_\Omega=\pm3.22\times 10^{-18}\,\mathrm{N}$ using DFT B3PW91/aug-cc-pDVZ, $F_z\rvert_\Omega=\pm 2.20\times 10^{-18}\,\mathrm{N}$ and $F_z\rvert_\Omega=\pm 2.33\times 10^{-18}\,\mathrm{N}$  using DFT CAM-B3LYP/aug-cc-pDVZ, $F_z\rvert_\Omega=\pm 1.35\times 10^{-18}\,\mathrm{N}$ and $F_z\rvert_\Omega=\pm 1.65\times 10^{-18}\,\mathrm{N}$ using DFT B3LYP/6-311G(d,p), and $F_z\rvert_\Omega=\pm 3.27\times 10^{-18}\,\mathrm{N}$ and $F_z\rvert_\Omega=\pm 3.60\times 10^{-18}\,\mathrm{N}$  using DFT B3LYP/6-311G++(2d,2p). These results are similar in magnitude and agree in sign, confirming that the essential features of our predictions are robust.}

For achiral eucalyptol with isotopic constitution ${}^{12}\mathrm{C}_{10}{}^1\mathrm{H}_{18}{}^{16}\mathrm{O}$, we obtained
\begin{align}
\frac{1}{2\pi\hbar}\left[\begin{matrix}
C & B & A
\end{matrix}\right]={}&\left[\begin{matrix}
1.02977 & 1.07159 & 1.53893
\end{matrix}\right]\,\mathrm{GHz}, \nonumber \\
\left[\begin{matrix}
\mu_{0c} & \mu_{0b} & \mu_{0a}
\end{matrix}\right]={}&\left[\begin{matrix}
0 & 4.92754 & -0.17782
\end{matrix}\right]\times10^{-30}\,\mathrm{C}\,\mathrm{m}, \nonumber \\
\left[\begin{matrix}
\alpha_{cc} & \alpha_{cb} & \alpha_{ca} \\
\alpha_{bc} & \alpha_{bb} & \alpha_{ba} \\
\alpha_{ac} & \alpha_{ab} & \alpha_{aa}
\end{matrix}\right](\omega^\prime)={}&\left[\begin{matrix}
1913.18 & 0 & 0 \\
0 & 1905.54 & 6.98 \\
0 & 6.98 & 2133.68
\end{matrix}\right]\times10^{-42}\,\mathrm{C}^2\,\mathrm{m}^2\,\mathrm{J}^{-1}, \nonumber \\
\left[\begin{matrix}
\alpha_{cc} & \alpha_{cb} & \alpha_{ca} \\
\alpha_{bc} & \alpha_{bb} & \alpha_{ba} \\
\alpha_{ac} & \alpha_{ab} & \alpha_{aa}
\end{matrix}\right](\omega)={}&\left[\begin{matrix}
1955.75 & 0 & 0 \\
0 & 1946.78 & 6.94 \\
0 & 6.94 & 2182.40
\end{matrix}\right]\times10^{-42}\,\mathrm{C}^2\,\mathrm{m}^2\,\mathrm{J}^{-1}. \nonumber
\end{align}
Note that the chirally sensitive products $\mu_{0c}\alpha_{ba}(\omega)$, $\mu_{0b}\alpha_{ac}(\omega)$, and $\mu_{0a}\alpha_{cb}(\omega)$ vanish, as they should.

For our isotopically chiral eucalyptol molecule (identical to achiral eucalyptol above but with \textcolor{black}{seven hydrogen atoms replaced by deuterium atoms as illustrated in Fig. \ref{Fig6}}), we obtained
\begin{align}
\frac{1}{2\pi\hbar}\left[\begin{matrix}
C & B & A
\end{matrix}\right]={}&\left[\begin{matrix}
0.95774 & 0.98410 & 1.40288
\end{matrix}\right]\,\mathrm{GHz}, \nonumber \\
\left[\begin{matrix}
\mu_{0c} & \mu_{0b} & \mu_{0a}
\end{matrix}\right]={}&\left[\begin{matrix}
-1.22940 & 4.75609 & -0.42478
\end{matrix}\right]\times10^{-30}\,\mathrm{C}\,\mathrm{m}, \nonumber \\
\left[\begin{matrix}
\alpha_{cc} & \alpha_{cb} & \alpha_{ca} \\
\alpha_{bc} & \alpha_{bb} & \alpha_{ba} \\
\alpha_{ac} & \alpha_{ab} & \alpha_{aa}
\end{matrix}\right](\omega^\prime)={}&\left[\begin{matrix}
1913.18 & 1.15 & -10.51 \\
1.15 & 1906.98 & 16.25 \\
-10.51 & 16.25 & 2132.24
\end{matrix}\right]\times10^{-42}\,\mathrm{C}^2\,\mathrm{m}^2\,\mathrm{J}^{-1}, \nonumber \\
\left[\begin{matrix}
\alpha_{cc} & \alpha_{cb} & \alpha_{ca} \\
\alpha_{bc} & \alpha_{bb} & \alpha_{ba} \\
\alpha_{ac} & \alpha_{ab} & \alpha_{aa}
\end{matrix}\right](\omega)={}&\left[\begin{matrix}
1955.67 & 1.47 & -10.76 \\
1.47 & 1948.32 & 16.53 \\
-10.76 & 16.53 & 2180.94
\end{matrix}\right]\times10^{-42}\,\mathrm{C}^2\,\mathrm{m}^2\,\mathrm{J}^{-1}. \nonumber
\end{align}
The properties of the opposite enantiomer can be obtained by flipping the sign of the dipole moment $\pmb{\mu}_0$.

\section*{Application to isotopically chiral molecules}
\label{Application to isotopically chiral molecules}
\begin{figure}[h!]
\centering
\includegraphics[width=0.5\linewidth]{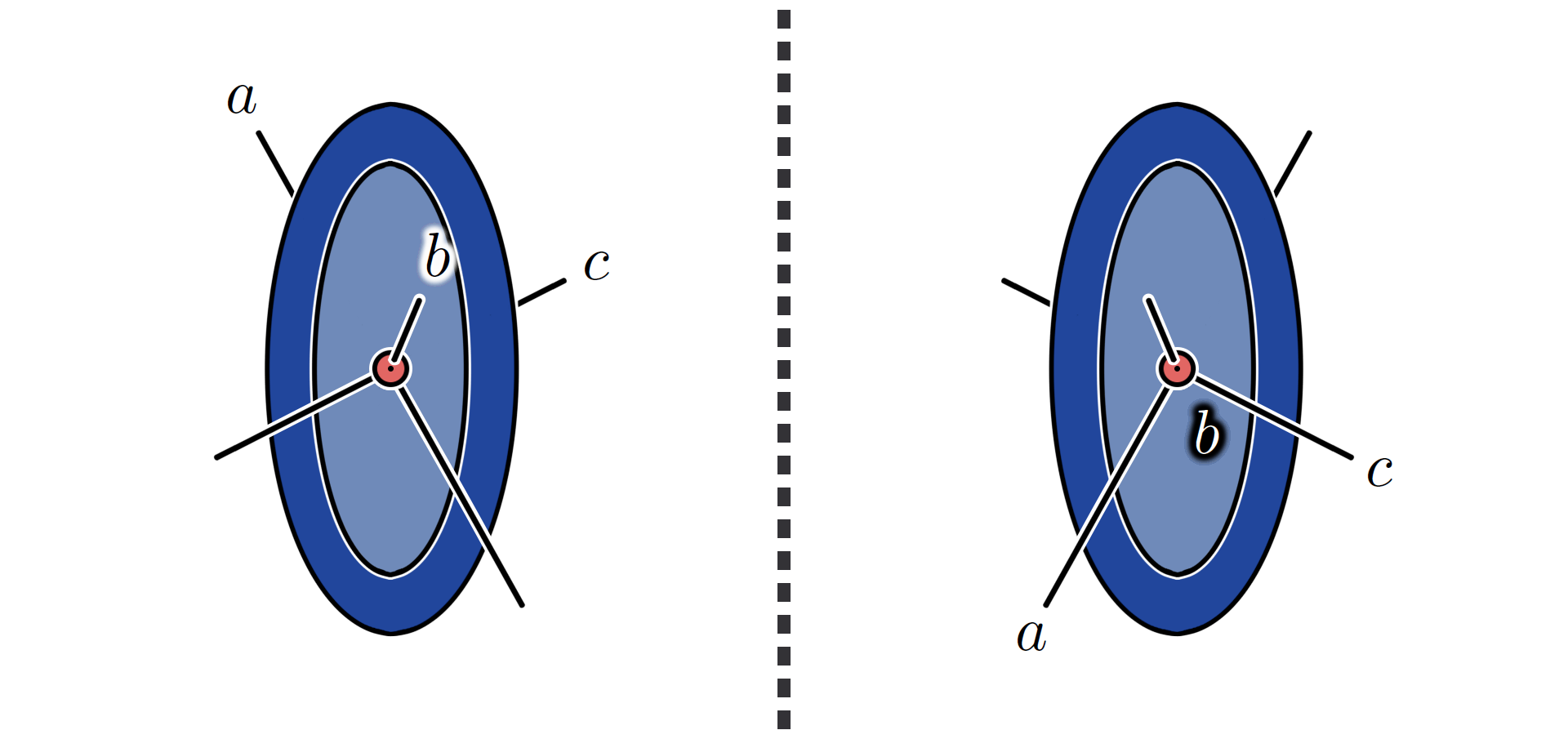}
\caption{\small \textcolor{black}{The (achiral) electronic properties of an isotopically chiral molecule can form a chiral geometry when taken together with the principal axes of inertia $c$, $b$, and $a$.}}
\label{Fig5}
\end{figure}

Isotopically chiral molecules constitute an interesting and potentially very important special case for our chiral optical force $\mathbf{F}$. The opposite enantiomers of an isotopically chiral molecule differ solely by the arrangement of their neutrons, making them notoriously difficult to \textcolor{black}{resolve} using existing methods \cite{Kimata97a}. As the electronic properties of an isotopically chiral molecule are essentially achiral (neglecting vibrational contributions), the strong-field limit given by Eq. \ref{StrongF} is essentially zero. The molecule's dipole moment $\pmb{\mu}_0$, traveling-wave \textcolor{black}{frequency} polarizability $\alpha(\omega^\prime)$, and standing-wave \textcolor{black}{frequency} polarizability $\alpha(\omega)$ can nevertheless form a chiral geometry embodying the molecule's chiral mass distribution when taken together the molecule's principal axes of inertia $c$, $b$, and $a$ as illustrated in Fig. \ref{Fig5}. If one or more of the products $\mu_{0c}\alpha_{ba}(\omega)$, $\mu_{0b}\alpha_{ac}(\omega)$, and $\mu_{0a}\alpha_{cb}(\omega)$ is non-zero, the weak-field limit given by Eq. \ref{PerturbativeF} gives non-trivial results, similar to an ordinary chiral molecule. Optima can be found between these extremes.

\begin{figure}[h!]
\centering
\includegraphics[width=0.5\linewidth]{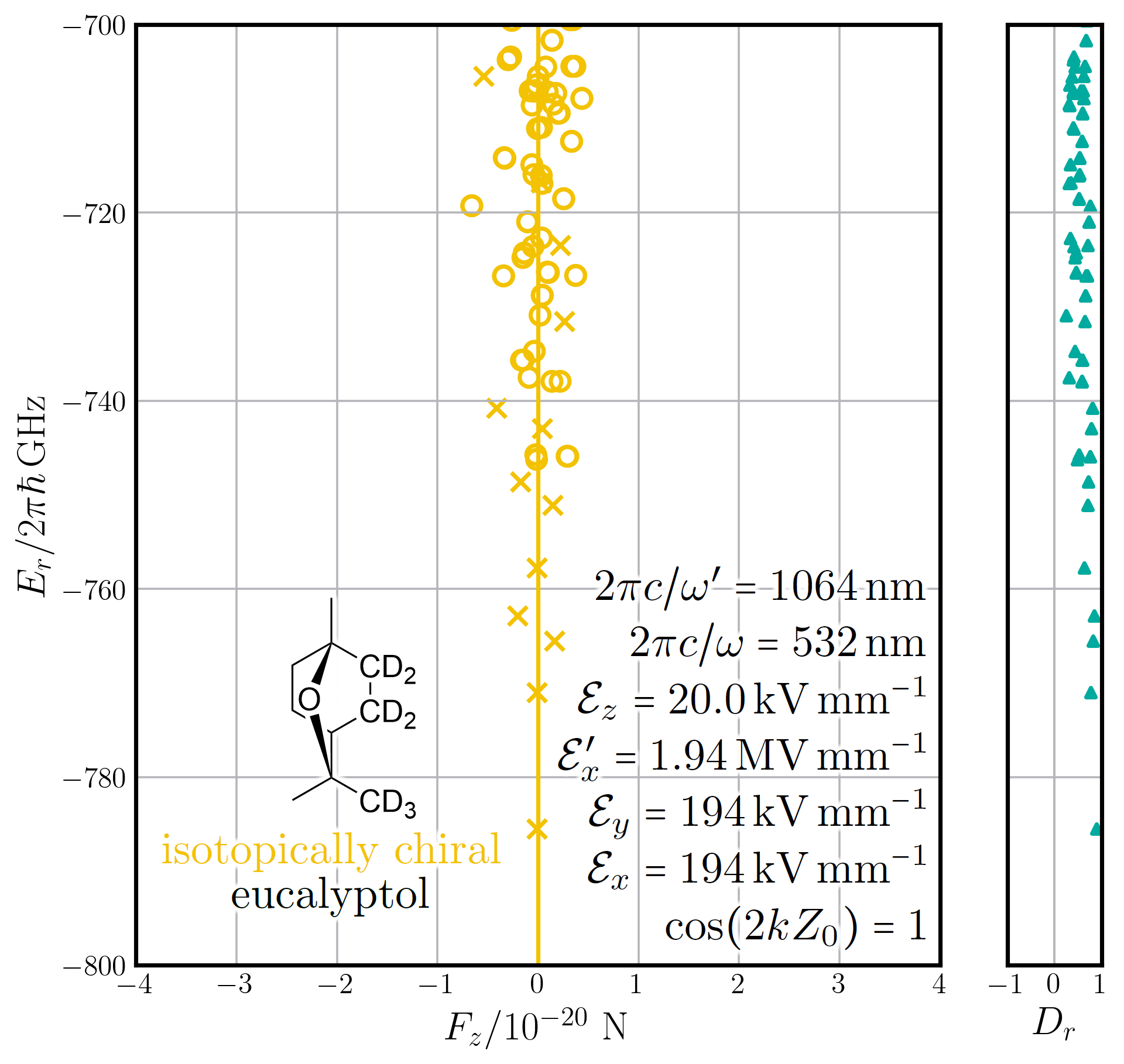} 
\caption{\small As in Figs. \ref{Fig2} and \ref{Fig4} but for a single enantiomer of isotopically chiral eucalyptol. Circles are non-degenerate.}
\label{Fig6}
\end{figure}

\textcolor{black}{Shown in Fig. \ref{Fig6} are results for an enantiomer of isotopically chiral eucalyptol calculated as described above. For the lowest energy rotational states, we approach the strong-field limit $F_z\rvert_\Omega=0$ given by Eq. \ref{StrongF} with the degree of orientation $D_r$ being close to $1$. As the rotational energy increases, the molecule's chiral mass distribution comes into play via the unperturbed rotational Hamiltonian $H^{(0)}$ and the force values reach appreciable magnitudes of $|\mathbf{F}|\sim 10^{-20}\,\mathrm{N}$ with a corresponding fall off in the degree of orientation $D_r$. The overall enantioselectivity is poor; $F_z$ has different signs for different rotational states for the single enantiomer considered here. Let us emphasize, however, that for each rotational state individually, $F_z$ has the same magnitude but opposite sign for the opposite enantiomer (not shown). Our chiral optical force might therefore prove useful for isotopically chiral molecules when combined with a suitable rotational state preparation scheme. In contrast, Eq. \ref{Fh} gives $\mathbf{F}^\prime\approx 0$ for isotopically chiral molecules, as $G^\prime(\omega)\approx 0$.}

\section*{Theoretical model of experiment}
\label{Theoretical model of experiment}
\textcolor{black}{In what follows, we show complicated functional dependencies explicitly to avoid any possible ambiguities.}

\subsection{Buffer gas molecular beam}
\textcolor{black}{A cryogenically cooled helium buffer gas beam of matrine enters an experimental chamber via a circular aperture of diameter $d_\mathrm{c}=1.00\,\mathrm{mm}$ located at $x=x_\mathrm{c}=-1.10\,\mathrm{m}$ with mean forward velocity in the range \cite{Hutzler12a, Samanta20a} 
\begin{align}
21.8\,\mathrm{m}\,\mathrm{s}^{-1}=\frac{3}{2}\sqrt{\frac{\pi k_\mathrm{B}T_\mathrm{c}}{2M}}\lesssim\overline{V}_\mathrm{f}\lesssim\sqrt{\frac{5k_\mathrm{B}T_\mathrm{c}}{m}}=204\,\mathrm{m}\,\mathrm{s}^{-1}, \label{Vfbar}
\end{align}
where $T_\mathrm{c}=4\,\mathrm{K}$ is the cell temperature, $M=4.12\times10^{-25}\,\mathrm{kg}$ is the mass of a matrine molecule, and $m=6.65\times10^{-27}\,\mathrm{kg}$ is the mass of a helium atom. We take the ratio of molecules to helium atoms in the beam to be $q\lesssim 1\times  10^{-2}$ and the experimental chamber to be evacuated to a residual pressure of $p\lesssim 10^{-8}\,\mathrm{mbar}$ at an ambient temperature of $T=300\,\mathrm{K}$. The lower limit in Eq. \ref{Vfbar} corresponds to the effusive regime, in which the helium stagnation number density in the cell satisfies $n_{\mathrm{c},\mathrm{He}}\lesssim1/2\sqrt{2}\sigma_{\mathrm{He}-\mathrm{He}}d_\mathrm{c}\sim 10^{14}\,\mathrm{cm}^{-3}$, assuming a helium-helium elastic collision cross section of $\sigma_{\mathrm{He}-\mathrm{He}}\sim 10^{-14}\,\mathrm{cm}^2$. The upper limit corresponds to the hydrodynamic regime, in which $n_{\mathrm{c},\mathrm{He}}\gtrsim 50/\sqrt{2}\sigma_{\mathrm{He}-\mathrm{He}}d_\mathrm{c}\sim 10^{16}\,\mathrm{cm}^{-3}$. We consider continuous beam operation.}

\textcolor{black}{The molecular properties quoted in what follows pertain to a single isotopologue of a single conformer of a single enantiomer with the understanding that the final results can be averaged with respect to isotopic abundances, conformational populations and enantiomeric purity as required.}

\textcolor{black}{We take the molecules emerging from the cell to occupy their vibronic ground states and a Maxwell-Boltzmann distribution of the unperturbed rotational energy eigenstates $|r\rangle^{(0)}$ with probabilities
\begin{align}
\int p_r(W)\,\mathrm{d}W={}&\frac{\mathrm{e}^{-E_r^{(0)}/k_\mathrm{B}T_\mathrm{c}}}{\sum_{r^\prime}\mathrm{e}^{-E_{r^\prime}^{(0)}/k_\mathrm{B}T_\mathrm{c}}}, \nonumber
\end{align}
where the $E_r^{(0)}$ are the associated rotational energy eigenvalues and $p_r(W)$ is the rotational state and translational energy probability density.}

\subsection{Collision rates}
\textcolor{black}{To gauge collision rates, let us focus here on the simple case of an effusive beam with a helium stagnation number density in the cell of $n_{\mathrm{c},\mathrm{He}}\sim 10^{14}\,\mathrm{cm}^{-3}$, adopting a hard spheres model of elastic collisions with velocity-independent cross sections. At a distance of $x_\mathrm{a}-x_\mathrm{c}=10\,\mathrm{cm}$ away from the cell, the mean collision rate of a molecule with helium atoms in the beam is \cite{Lubman82a}
\begin{align}
\overline{\Gamma}_\mathrm{m-He}(x_\mathrm{a})={}&\frac{n_{\mathrm{c},\mathrm{He}}d_\mathrm{c}^2\sigma_{\mathrm{m-He}}}{16(x_\mathrm{a}-x_\mathrm{c})^2}\sqrt{\frac{8k_\mathrm{B}T_\mathrm{c}}{\pi M}}\left[\frac{2(M^2+m^2)+3Mm-(\sqrt{M}+\sqrt{m})(M+m)^{3/2}}{(M+m)^{3/2}\sqrt{m}}\right]\sim 10^{-1}\,\mathrm{s}^{-1} \label{GammamHebar}
\end{align}
and the mean collision rate of a molecule in the beam with other molecules in the beam is \cite{Lubman82a}
\begin{align}
\overline{\Gamma}_\mathrm{m-m}(x_\mathrm{a})={}&\frac{qn_{\mathrm{c},\mathrm{He}}d_\mathrm{c}^2\sigma_{\mathrm{m-m}}}{16(x_\mathrm{a}-x_\mathrm{c})^2}\sqrt{\frac{8k_\mathrm{B}T_\mathrm{c}}{\pi M}}\left(\frac{7\sqrt{2}-8}{4}\right)\sim10^{-4}\,\mathrm{s}^{-1}, \label{Gammammbar}
\end{align}
assuming a molecule-helium elastic collision cross section of $\sigma_\mathrm{m-He}\sim 10^{-14}\,\mathrm{cm}^2$ \cite{Hutzler12a} and a molecule-molecule elastic collision cross section of $\sigma_\mathrm{m-m}\sim 10^{-13}\,\mathrm{cm}^2$ \cite{Patterson10a}. Note that $\overline{\Gamma}_\mathrm{m-He}(x)$ and $\overline{\Gamma}_\mathrm{m-m}(x)$ reduce quadratically with increasing distance away from the cell. The mean collision rate of a molecule in the beam with residual helium atoms in the experimental chamber is
\begin{align}
\overline{\Gamma}_\mathrm{m-res}={}&\frac{\sigma_{\mathrm{m-He}}p}{k_\mathrm{B}T}\sqrt{\frac{8k_\mathrm{B}(M T+mT_\mathrm{c})}{\pi Mm}}\sim 10^{-1}\,\mathrm{s}^{-1}, \label{Gammamresbar}
\end{align}
again assuming $\sigma_\mathrm{m-He}\sim 10^{-14}\,\mathrm{cm}^2$ \cite{Hutzler12a}. Eqs. \ref{GammamHebar}-\ref{Gammamresbar} show that a molecule in the beam is most likely to collide with a residual helium atom in the experimental chamber ($\overline{\Gamma}_\mathrm{m-res}>\overline{\Gamma}_\mathrm{m-He}(x)\gg \overline{\Gamma}_\mathrm{m-m}(x)$ for $x>x_\mathrm{a}$).}

\textcolor{black}{As the molecules are internally cold and essentially collision free as they traverse the remainder of the experimental chamber ($\overline{\Gamma}_\mathrm{m-res}(x_\mathrm{PECD}-x_\mathrm{a})/(2\hbar W_r/M)^{1/2}\ll 1$), we henceforth adopt a single-molecule description with the molecule's center-of-mass motion described quantum mechanically in terms of matter waves \cite{Arndt99a, Brand21a}, letting $\mathbf{R}_0\rightarrow\mathbf{R}=Z\hat{\mathbf{z}}+Y\hat{\mathbf{y}}+X\hat{\mathbf{x}}$ denote the molecule's unfrozen position and $\hbar W$ denote the molecule's (initial) translational energy, $W$ being the associated angular frequency.} 

\textcolor{black}{We first analyze the experiment for a single unperturbed rotational state $|r\rangle^{(0)}$ and translational energy $\hbar W$, then average over the rotational state and translational energy probability density $p_r(W)$.}

\subsection{Collimating aperture}
\textcolor{black}{The beam passes first through a circular collimating aperture of diameter $d_\mathrm{a}=1.00\,\mu\mathrm{m}$ located at $x=x_\mathrm{a}=-1.00\,\mathrm{m}$, generating transverse matter-wave coherence downstream \cite{Brand21a}.}

\subsection{Static field}
\textcolor{black}{The static field $\mathbf{E}_0(z,y,x)$ is generated by four cylindrical electrodes of length $l_\mathrm{e}=20\,\mathrm{cm}$ and diameter $d_\mathrm{e}=2\,\mathrm{mm}$ running parallel to the $x$ axis in a quadrupole geometry of width $w_\mathrm{e}=1\,\mathrm{mm}$ centered on $x=x_\mathrm{e}=0$. The electrodes are connected in pairs on either side of the $y$-$x$ plane to the terminals of a high-voltage dc power supply. Note that this geometry permits access for the laser pulses described below.}

\textcolor{black}{As the transverse dc-Stark forces due to the static field $\mathbf{E}_0(z,y,x)$ are weaker than our chiral optical force $\mathbf{F}$ ($|\pmb{\mu}_0\rvert_\Omega\cdot\partial_Z\mathbf{E}_0(Z,Y,X)|,|\pmb{\mu}_0\rvert_\Omega\cdot\partial_Y\mathbf{E}_0(Z,Y,X) |\ll|k\mathcal{E}_y\mathcal{E}_x\alpha_{yx}(\omega)\rvert_\Omega|$ for $|Z|\lesssim w_\mathrm{g}$ and $|Y|\lesssim h_\mathrm{g}$), we consider the simplifed form
\begin{align}
\mathbf{E}_0(x)={}&f_0(x)\mathcal{E}_z\hat{\mathbf{z}}, \label{StaticEGeneral}
\end{align}
where $f_0(x)$ with $f_0(\mp\infty)=0$ and $f_0(x_\mathrm{e})=1$ accounts for the field's spatial variation and $\mathcal{E}_z=20.0\,\mathrm{kV}\,\mathrm{mm}^{-1}$ is the field's maximum strength, assuming a positive electrode polarity. Note that Eq. \ref{StaticEGeneral} is a generalization of Eq. \ref{StaticE}.}

\textcolor{black}{When the molecule enters the static field $\mathbf{E}_0(x)$, the unperturbed rotational energy eigenvalue  $E_r^{(0)}$ and associated energy eigenstate $|r\rangle^{(0)}$ change adiabatically into the dc-Stark-shifted rotational energy eigenvalue $E_r^{(\mu)}(X)$ and associated energy eigenstate $|r(X)\rangle^{(\mu)}$, obtained by solving the rotational energy eigenvalue equation
\begin{align}
H^{(\mu)}(X)|r(X)\rangle^{(\mu)}={}&E_r^{(\mu)}(X)|r(X)\rangle^{(\mu)}, \nonumber
\end{align}
where 
\begin{align}
H^{(\mu)}(X)={}&-f_0(X)\mathcal{E}_z\mu_{0z}+H^{(0)} \nonumber
\end{align}
is the rotational Hamiltonian in the presence of $\mathbf{E}_0(x)$. When the molecule exits $\mathbf{E}_0(x)$, $E_r^{(\mu)}(X)$ and $|r(X)\rangle^{(\mu)}$ change adiabatically back into $E_r^{(0)}$ and $|r\rangle^{(0)}$. Thus, $E_r^{(\mu)}(\mp\infty)=E_r^{(0)}$ and $|r(\mp\infty)\rangle^{(\mu)}=|r\rangle^{(0)}$. As $\mathbf{E}_0(x)$ varies slowly in $x$ relative to the matter waves ($\partial_X\psi_r(Z,Y,X,t;W)\partial_X|r(X)\rangle^{(\mu)},\psi_r(Z,Y,X,t;W)\partial_X^2|r(X)\rangle^{(\mu)}\ll\partial_X^2\psi_r(Z,Y,X,t;W)|r(X)\rangle^{(\mu)}$), we take the molecule's translational degrees of freedom to be governed by the Schr\"{o}dinger equation in the form
\begin{align}
\mathrm{i}\hbar\partial_t\psi_r(Z,Y,X,t;W)={}&\left\{-\frac{\hbar^2}{2M}(\partial_Z^2+\partial_Y^2+\partial_X^2)+\left[E^{(\mu)}_r(X)-E_r^{(0)}\right]\right\}\psi_r(Z,Y,X,t;W), \ \ \ \ \label{SchroedingerEquation}
\end{align}
where $\psi_r(Z,Y,X,t;W)$ is the molecule's translational wavefunction and the dc Stark shift $E^{(\mu)}_r(X)-E_r^{(0)}$ serves as an effective translational potential energy. To help us solve Eq. \ref{SchroedingerEquation} in the WKB approximation, we identify the local angular wavenumber
\begin{align}
K_r(X;W)={}&\sqrt{\frac{2M}{\hbar^2}\left\{\hbar W-\left[E^{(\mu)}_r(X)-E_r^{(0)}\right]\right\}}. \label{LocalK}
\end{align}
When the molecule enters and then exits $\mathbf{E}_0(x)$, $K_r(X;W)$ changes from $K(-\infty;W)$ to $K_r(x_\mathrm{e};W)$ and then returns to $K(+\infty;W)=K(-\infty;W)$. Eq. \ref{LocalK} gives $K(\mp\infty;W)=3.91\times 10^{11}\,\mathrm{rad}\,\mathrm{m}^{-1}$ and $K_\Omega(x_\mathrm{e};W)/K(\mp\infty;W)=1.07$ for a notional molecule with a translational energy of $\hbar W=12.9\,\mathrm{meV}$ ($(2\hbar W/M)^{1/2}=100\,\mathrm{m}\,\mathrm{s}^{-1}$) and perfect orientation of the dipole moment $\pmb{\mu}_0$ along $\mathbf{E}_0(x)$ ($E^{(\mu)}_\Omega(X)-E_\Omega^{(0)}=-f_0(X)\mathcal{E}_z|\pmb{\mu}_0|$), showing that a local de Broglie wavelength initially comparable to $2\pi/K(\mp\infty;W)=16.1\,\mathrm{pm}$ can vary by several percent as the molecule enters and then exits $\mathbf{E}_0(x)$.}

\subsection{Mechanical grating}
\textcolor{black}{A thin mechanical grating located at $x=x_\mathrm{g}=0$ mitigates the $\cos(2kZ)$ modulation of our chiral optical force $\mathbf{F}$ by only transmitting regions of the matter waves for which $\cos(2kZ)\ge0$. We consider $2N_\mathrm{g}+1=11$ slits of height $h_\mathrm{g}=1.00\,\mu\mathrm{m}$ and width $26.6\,\mathrm{nm}=\pi/10k\le d_\mathrm{g}\le\pi/2k=133\,\mathrm{nm}$ arranged with a matching period of $\pi/k=266\,\mathrm{nm}$ along $z$, giving a total width of $2.69\,\mu\mathrm{m}\le w_\mathrm{g}=2\pi N_\mathrm{g}/k+d_\mathrm{g}\le 2.79\,\mu\mathrm{m}$. The effect of van der Waals interactions between the molecules and the mechanical grating can be modeled as an effective reduction of the slit width $d_\mathrm{g}$ \cite{Arndt99a}.}

\textcolor{black}{We take the transverse matter-wave coherence length at the mechanical grating to be
\begin{align}
\xi_r(x_\mathrm{g};W)={}&\frac{2.44\pi}{d_\mathrm{a}}\int_{x_\mathrm{a}}^{x_\mathrm{g}}\frac{\mathrm{d}X^\prime}{K_r(X^\prime;W)},   \label{TransverseCoherenceLength}
\end{align}
assuming no transverse coherence at the collimating aperture. Eq. \ref{TransverseCoherenceLength} gives $\xi(x_\mathrm{g};W)=19\,\mu\mathrm{m}$ for a notional molecule with a translational energy of $\hbar W=12.9\,\mathrm{meV}$ ($(2\hbar W/M)^{1/2}=100\,\mathrm{m}\,\mathrm{s}^{-1}$) and perfect orientation of the dipole moment $\pmb{\mu}_0$ along the static field $\mathbf{E}_0(x)$ ($E^{(\mu)}_\Omega(X)-E_\Omega^{(0)}=-f_0(X)\mathcal{E}_z|\pmb{\mu}_0|$), showing that the matter waves have excellent transverse coherence at the mechanical grating.}

\textcolor{black}{As the transverse matter-wave coherence length is larger than the mechanical grating's slit height and total width ($\xi_r(x_\mathrm{g};W)\gg w_\mathrm{g},h_\mathrm{g}$), we take the molecule's translational wavefunction downstream from the mechanical grating and prior to the pulsed illumination described below to be
\begin{align}
\psi_r(Z,Y,X,t;W)={}&\frac{1}{4\pi^2}\iiiint\sqrt{\frac{K_{x,r}(x_\mathrm{g},K_z,K_y;W)}{K_{x,r}(X,K_z,K_y;W)}}\psi_r(Z^\prime,Y^\prime,x_\mathrm{g},t;W) \nonumber \\
{}&\times\mathrm{e}^{\mathrm{i}\left[K_z(Z-Z^\prime)+K_y(Y-Y^\prime)+\int_{x_\mathrm{g}}^XK_{x,r}(X^\prime,K_z,K_y;W)\,\mathrm{d}X^\prime\right]}\,\mathrm{d}Z^\prime\mathrm{d}Y^\prime\mathrm{d}K_z\mathrm{d}K_y \ \ \ \ (X\ge x_\mathrm{g},t<t_\mathrm{p}) \label{WavefunctionBeforePulses}
\end{align}
with
\begin{align}
\psi_r(Z,Y,x_\mathrm{g},t;W)={}&\frac{1}{\sqrt{(2N_\mathrm{g}+1)h_\mathrm{g}d_\mathrm{g}}}\sum_{s=-{N_\mathrm{g}}}^{N_\mathrm{g}}\mathrm{rect}\left(\frac{Z-s\pi/k}{d_\mathrm{g}}\right)\mathrm{rect}\left(\frac{Y}{h_\mathrm{g}}\right)\mathrm{e}^{-\mathrm{i}W t} \label{GratingWavefunction} \\
K_{x,r}(X,K_z,K_y;W)={}&\sqrt{K_r^2(X;W)-K_z^2-K_y^2}, 
\end{align}
where $\psi_r(Z,Y,x_\mathrm{g},t;W)$ is the molecule's translational wavefunction in the plane of the mechanical grating, featuring transverse wavevector spreads of $9.45\times 10^7\,\mathrm{rad}\,\mathrm{m}^{-1}\le\Delta K_z=4\pi/d_\mathrm{g}\le 4.72\times 10^8\,\mathrm{rad}\,\mathrm{m}^{-1}$ and $\Delta K_y=4\pi/h_\mathrm{g}=1.26\times 10^7\,\mathrm{rad}\,\mathrm{m}^{-1}$, and $K_{x,r}(X,K_z,K_y;W)$ is the local longitudinal wavevector for given transverse wavevectors $K_z$ and $K_y$. Eq. \ref{WavefunctionBeforePulses} is the solution of Eq. \ref{SchroedingerEquation} in the WKB approximation, obtained using the angular spectrum method.}

\textcolor{black}{The $\cos(2kZ)$ modulation could also be mitigated using spatially varying matter-wave depletion via optical absorption \cite{Walter16a}.}

\subsection{Spectral filtering (velocity selection)}
\textcolor{black}{Together, the collimating aperture and mechanical grating provide longitudinal matter-wave coherence via gravitationally mediated spectral filtering (velocity selection) \cite{Brand21a}. We account here for this important effect heuristically; a more rigorous description would follow from the inclusion of the molecule's gravitational potential energy in Eq. \ref{SchroedingerEquation}. On the basis of kinematical arguments, we take the fractional width of the rotational state and translational energy probability density $p_r(W)$ downstream from the mechanical grating to be
\begin{align}
\frac{\Delta W_r}{\overline{W}_r}={}&\frac{2\hbar^2 h_\mathrm{g}}{g M^2 K_r^2(-\infty;\overline{W}_r)}\left[\int_{x_\mathrm{a}}^{x_\mathrm{g}}\frac{\mathrm{d}X^\prime}{K_r(X^\prime;\overline{W}_r)}\int_{x_\mathrm{a}}^{x_\mathrm{g}}\frac{\mathrm{d}X^{\prime\prime}}{K_r^3(X^{\prime\prime};\overline{W}_r)}\right]^{-1}, \label{EnergyWidth}
\end{align}
assuming horizontal translation at the collimating aperture, where $\overline{W}_r$ is the mean translational energy (dictated by the downwards displacement of the mechanical grating relative to the collimating aperture) and $g=9.81\,\mathrm{m}\,\mathrm{s}^{-2}$ is the gravitational field strength. Eq. \ref{EnergyWidth} gives $\Delta W_r/\overline{W}_r\sim 10^{-3}$ for a mean translational energy of $\hbar\overline{W}_r=12.9\,\mathrm{meV}$ $(2\hbar\overline{W}_r/M)^{1/2}=100\,\mathrm{m}\,\mathrm{s}^{-1}$), showing that the matter waves have excellent longitudinal coherence downstream from the mechanical grating. Note that different rotational states correspond to different values of $\overline{W}_r$ due to the static field $\mathbf{E}_0(x)$, providing an opportunity for effective rotational state selectivity via the introduction of additional apertures.}

\subsection{Laser pulses}
\textcolor{black}{The travelling-wave field $\tilde{\mathbf{E}}^\prime(z,y,x,t)$ and standing-wave field $\tilde{\mathbf{E}}(z,y,x,t)$ are realized in pulsed form using a commercial Nd:YAG laser at the fundamental wavelength of $2\pi c/\omega^\prime=1064\,\mathrm{nm}$ and second-harmonic of $2\pi c/\omega=532\,\mathrm{nm}$. The laser pulses have Gaussian forms with spatial widths of $w_z^\prime=250\,\mu\mathrm{m}$ and $w_x^\prime=100\,\mu\mathrm{m}$ for the travelling wave, $w_y=250\,\mu\mathrm{m}$ and $w_x=100\,\mu\mathrm{m}$ for the standing wave, and a common temporal width of $\tau_\mathrm{p}=10.0\,\mathrm{ns}$. We consider a single focal point located at $x=x_\mathrm{p}$ at time $t=t_\mathrm{p}=0$ with the optimal value of $x_\mathrm{p}$ being dictated by the Talbot effect as described below. The lin$\perp$lin configuration is formed by retroreflecting the second-harmonic laser pulse from a mirror combined with a quarter-wave plate.}

\textcolor{black}{As the laser pulses are paraxial ($k^\prime w_z^\prime,k^\prime w_x^\prime,kw_y,kw_x\gg1$), have larger transverse extents than the matter waves ($w^\prime_z\gg w_\mathrm{g}$, $w_y\gg h_\mathrm{g}$), diffract little over the extent of the matter waves ($h_\mathrm{g}/k^\prime w_z^{\prime 2},h_\mathrm{g}/k^\prime w_x^{\prime 2},w_\mathrm{g}/kw_y^2,w_\mathrm{g}/kw_x^2\ll 1$), have negligible retardation  over the extent of the matter waves ($w_\mathrm{g},h_\mathrm{g}\ll c\tau_\mathrm{p}$), and give transverse envelope forces smaller than our chiral optical force ($|\partial_Z[\tilde{E}_\beta^\prime(Z,Y,X,t)\tilde{E}_\alpha^{\prime\ast}(Z,Y,X,t)]\alpha_{\beta\alpha}(\omega^\prime)\rvert_\Omega/4|,\partial_Y[\tilde{E}_\beta(Z,Y,X,t)\tilde{E}_\alpha^\ast(Z,Y,X,t)]\alpha_{\beta\alpha}(\omega)\rvert_\Omega/4|\ll|k\mathcal{E}_y\mathcal{E}_x\alpha_{yx}(\omega)\rvert_\Omega|$), we henceforth consider the simplifed forms
\begin{align}
\tilde{\mathbf{E}}^\prime(y,x,t)={}&f_{\omega^\prime}(x,t)\mathcal{E}_x^\prime\mathrm{e}^{\mathrm{i}(k^\prime y-\omega^\prime t)}\hat{\mathbf{x}}, \label{ComplexEPrimeGeneral} \\
\tilde{\mathbf{E}}(z,x,t)={}&f_\omega(x,t)\left[\mathcal{E}_y\mathrm{e}^{\mathrm{i}(kz-\omega t)}\hat{\mathbf{y}}+\mathrm{i}\mathcal{E}_x\mathrm{e}^{\mathrm{i}(-kz-\omega t)}\hat{\mathbf{x}}\right] \label{ComplexEGeneral}
\end{align}
with
\begin{align}
f_{\omega^\prime}(x,t)={}&\mathrm{e}^{-(x-x_\mathrm{p})^2/w_x^{\prime2}}\mathrm{e}^{-(t-t_\mathrm{p})^2/\tau_\mathrm{p}^2}, \nonumber \\
f_\omega(x,t)={}&\mathrm{e}^{-(x-x_\mathrm{p})^2/w_x^2}\mathrm{e}^{-(t-t_\mathrm{p})^2/\tau_\mathrm{p}^2}, \nonumber
\end{align}  
where $f_{\omega^\prime}(x,t)$ accounts for the spatiotemporal variation of the travelling-wave field $\tilde{\mathbf{E}}^\prime(y,x,t)$ with a peak field strength of $\mathcal{E}_x^\prime=1.94\,\mathrm{MV}\,\mathrm{mm}^{-1}$ and $f_\omega(x,t)$ accounts for the spatiotemporal variation of the standing-wave field $\tilde{\mathbf{E}}(z,x,t)$ with peak field strengths of $\mathcal{E}_y=194\,\mathrm{kV}\,\mathrm{mm}^{-1}\approx\mathcal{E}_x$; small losses due to the quarter-wave plate and mirror can be accounted for by taking $\mathcal{E}_x\lesssim\mathcal{E}_y$. Note that Eqs. \ref{ComplexEPrimeGeneral} and \ref{ComplexEGeneral} are generalizations of Eqs. \ref{ComplexEPrime} and \ref{ComplexE}.}

\textcolor{black}{As the molecule moves only slightly during the illumination ($\hbar\Delta K_z\tau_\mathrm{p}/M\ll2\pi/k$; $\hbar K_{r,x}(X,K_z,K_y;W)\tau_\mathrm{p}/M\ll w_x^\prime,w_x$), we model the effect of the laser pulses by taking
\begin{align}
\psi_r(Z,Y,X,t_\mathrm{p};W)\rightarrow \psi^\prime_r(Z,Y,X,t_\mathrm{p};W)={}&I_r(Z,X,t_\mathrm{p})\psi_r(Z,Y,X,t_\mathrm{p};W), \nonumber
\end{align}
where $I_r(Z,X,t_\mathrm{p})$ is a phase modulation imprinted upon the molecule's translational wavefunction $\psi_r(Z,Y,X,t_\mathrm{p};W)$ around time $t=t_\mathrm{p}$, giving a modified translational wavefunction $\psi^\prime_r(Z,Y,X,t_\mathrm{p};W)$. As the molecule can rotate faster than the pulse's temporal width ($6\pi\hbar/J(J+1)(C+B+A)\ll\tau_\mathrm{p}$ for $J\in\{1,2,\dots\}$), we assume adiabatic following and consider
\begin{align}
I_r(Z,X,t_\mathrm{p})={}&\mathrm{e}^{-\mathrm{i}\int [E_r(Z,X,t^\prime)-E_r^{(\mu)}(X)]\,\mathrm{d}t^\prime/\hbar}, \label{IInitial}
\end{align}
where the instantaneous rotational energy eigenvalue $E_r(Z,X,t)$ and associated energy eigenstate $|r(Z,X,t^\prime)\rangle$ are obtained by solving the instantaneous rotational energy eigenvalue equation
\begin{align}
H(Z,X,t^\prime)|r(Z,X,t^\prime)\rangle={}&E_r(Z,X,t^\prime)|r(Z,X,t^\prime)\rangle, \label{HInstantan}
\end{align}
the instantaneous rotational Hamiltonian being
\begin{align}
H(Z,X,t^\prime)={}&-f_0(X)\mathcal{E}_z\mu_{0z}-f_{\omega^\prime}^2(X,t^\prime)\frac{1}{4}\mathcal{E}_x^{\prime 2}\alpha_{xx}(\omega^\prime) \nonumber \\
{}&-f_\omega^2(X,t^\prime)\bigg[\frac{1}{4}\mathcal{E}_y^2\alpha_{yy}(\omega)+\frac{1}{4}\mathcal{E}_x^2\alpha_{xx}(\omega)+\frac{1}{2}\mathcal{E}_y\mathcal{E}_x\alpha_{yx}(\omega)\sin(2kZ)\bigg]+H^{(0)}. \label{Hrotrans}
\end{align}
We consider the form
\begin{align}
E_r(Z,X,t^\prime)={}&-\Delta_r(X,t^\prime)-\frac{1}{2}\mathcal{E}_y\mathcal{E}_x\chi_r(X,t^\prime)\sin(2kZ)-\frac{1}{2}\mathcal{E}_y\mathcal{E}_x\sum_{n=1}^\infty\frac{1}{(n+1)}\Upsilon_r^{(n)}(X,t^\prime)\sin^{n+1}(2kZ). \label{ErGeneral}
\end{align}
Note that Eqs. \ref{HInstantan}, \ref{Hrotrans}, and \ref{ErGeneral} are generalizations of Eqs. \ref{HrEr}, \ref{H}, and \ref{Er}.}

\textcolor{black}{We take the molecule's modified translational wavefunction in the half space downstream from the mechanical grating and after the pulsed illumination to be
\begin{align}
\psi^\prime_r(Z,Y,X,t;W)={}&\frac{1}{4\pi^2}\iiiint\sqrt{\frac{K_{x,r}(x_\mathrm{g},K_z,K_y;W)}{K_{x,r}(X,K_z,K_y;W)}}I_r\left(Z,X,t;K_z,K_y,W\right)  \psi_r(Z^\prime,Y^\prime,x_\mathrm{g},t;W)    \nonumber \\
{}&\times  \mathrm{e}^{\mathrm{i}\left[K_z(Z-Z^\prime)+K_y(Y-Y^\prime)+\int_{x_\mathrm{g}}^XK_{x,r}(X^\prime,K_z,K_y;W)\,\mathrm{d}X^\prime\right]}\,\mathrm{d}Z^\prime\mathrm{d}Y^\prime\mathrm{d}K_z\mathrm{d}K_y \ \ \ \ (X\ge x_\mathrm{g},t>t_\mathrm{p}) \label{PsiPrimeGeneral}
\end{align}
with
\begin{align}
I_r\left(Z,X,t;K_z,K_y,W\right)={}&\frac{1}{2\pi}\mathrm{e}^{\mathrm{i}\int\{E_r^{(\mu)}[X_{\mathrm{cl},r}(X,K_z,K_y,t;W)]+\Delta_r[X_{\mathrm{cl},r}(X,K_z,K_y,t;W),t^\prime]\}\,\mathrm{d}t^\prime/\hbar} \nonumber \\
{}&\times\iint\mathrm{e}^{\mathrm{i}K_z^\prime(Z-Z^\prime)}\mathrm{e}^{\mathrm{i}\mathcal{E}_y\mathcal{E}_x\int\chi_r\left[X_{\mathrm{cl},r}(X,K_z,K_y,t;W),t^\prime\right]\,\mathrm{d}t^\prime\sin(2kZ^\prime)/2\hbar} \nonumber \\
{}&\times\mathrm{e}^{\mathrm{i}\mathcal{E}_y\mathcal{E}_x\sum_{n=1}^\infty\int\Upsilon_r^{(n)}\left[X_{\mathrm{cl},r}(X,K_z,K_y,t;W),t^\prime\right]\,\mathrm{d}t^\prime\sin^{n+1}(2kZ^\prime)/2(n+1)\hbar} \nonumber \\
{}&\times\mathrm{e}^{-\mathrm{i}\hbar K_z^\prime K_z(t-t_\mathrm{p})/M}\mathrm{e}^{-\mathrm{i}\hbar K_z^{\prime2}(t-t_\mathrm{p})/2M}\,\mathrm{d}Z^\prime\mathrm{d}K_z^\prime \label{Ir} \\
X_{\mathrm{cl},r}(X,K_z,K_y,t;W)={}&X-\frac{\hbar}{M}\int_{t_\mathrm{p}}^tK_{x,r}\left[X_{\mathrm{cl},r}(X,K_z,K_y,t^{\prime\prime};W);W\right]\,\mathrm{d}t^{\prime\prime}, \nonumber
\end{align}
where $X_{\mathrm{cl},r}(X,K_z,K_y,t;W)$ is a classical trajectory that accounts for retardation. Eq. \ref{Ir} is the solution of
\begin{align}
\mathrm{i}\hbar\partial_tI_r(Z,X,t;K_z,K_y,W)={}&-\frac{\mathrm{i}\hbar^2K_z}{M}\partial_ZI_r(Z,X,t;K_z,K_y,W)-\frac{\hbar^2}{2M}\partial_Z^2I_r(Z,X,t;K_z,K_y,W) \nonumber \\
{}&-\frac{\mathrm{i}\hbar^2K_{x,r}(X,K_z,K_y;W)}{M}\partial_XI_r(Z,X,t;K_z,K_y,W) \label{SchroedingerEnvelope}
\end{align}
that reduces to Eq. \ref{IInitial} at time $t=t_\mathrm{p}$, obtained using the method of characteristic curves. Eq. \ref{SchroedingerEnvelope} is itself obtained by substituting Eq. \ref{PsiPrimeGeneral} into Eq. \ref{SchroedingerEquation} then making the WKB approximation and the slowly varying envelope approximation with respect to the longitudinal variation of the imprinted phase modulation ($\partial_X^2 I_r(Z,X,t;K_z,K_y,W)\ll K_{x,r}(X,K_z,K_y;W)\partial_X I_r(Z,X,t;K_z,K_y,W)$).}

\textcolor{black}{As the angular spread of wavevectors is small ($\Delta K_z+100\hbar k,\Delta K_y,\ll K_{r}(X;W)$), the $\Upsilon^{(n)}_r$ corrections usually converge rapidly ($\Upsilon^{(n)}_r\approx 0$ for $n\in\{2,3,\dots\}$), and the location of the detection plane described below approaches the Fraunhofer domain ($w_\mathrm{g}^2+h_\mathrm{g}^2\ll 2\int_{x_\mathrm{g}}^{X}\mathrm{d}X^\prime/K_r(X^\prime;W)$ for $X\sim x_\mathrm{PECD}$) \cite{Brand21a}, we approximate Eq. \ref{PsiPrimeGeneral} in the vicinity of the detection plane by
\begin{align}
\psi^\prime_r(Z,Y,X,t;W)={}&\psi_r^{(0)}(X;W)\mathrm{e}^{\mathrm{i}\Phi_r(Z,Y,X,t;W)}\sum_{w=-\infty}^\infty\sum_{v=-\infty}^\infty\mathrm{J}_w\left[R_r(X,t;W)\right]\mathrm{J}_v\left[Q_r(X,t;W)\right]\mathrm{i}^{-v}\mathrm{e}^{\mathrm{i}\eta_{w,v;r}(Z,X,t;W)} \nonumber \\
{}&\times\mathrm{sinc}\left\{\left[\mathcal{Z}_r(Z,X;W)-\gamma_r(X,t;W)(w+2v)\right]kd_\mathrm{g}\right\}\mathrm{sinc}\left[\mathcal{Y}_r(Y,X;W)kh_\mathrm{g}\right] \nonumber \\
{}&\times\frac{\sin\left\{(2N_\mathrm{g}+1)\pi\left[\mathcal{Z}_r(Z,X;W)-\gamma_r(X,t;W)(w+2v)\right]\right\}}{\sin\left\{\pi\left[\mathcal{Z}_r(Z,X;W)-\gamma_r(X,t;W)(w+2v)\right]\right\}} \ \ \ \ (X\sim x_\mathrm{PECD},t>t_\mathrm{p}) \label{PsiPrimeFraunhofer}
\end{align}
with
\begin{align}
\psi_r^{(0)}(X;W)={}&\frac{M}{2\pi \hbar\Delta t_{\mathrm{g},r}(X;W)}\sqrt{\frac{K_r(x_\mathrm{g};W)h_\mathrm{g}d_\mathrm{g}}{(2N_\mathrm{g}+1)K_r(X;W)}}, \nonumber \\
\Phi_r(Z,Y,X,t;W)={}&-\frac{\pi}{2}+\int_{x_\mathrm{g}}^X K_r(X^\prime;W)\,\mathrm{d}X^\prime-Wt+\frac{2\hbar[\mathcal{Z}_r^2(Z,X;W)+\mathcal{Y}_r^2(Y,X;W)]\Delta t_{\mathrm{g},r}(X;W)k^2}{M} \nonumber \\
{}&+\frac{1}{\hbar}\int\left\{E_r^{(\mu)}[X_{\mathrm{cl},r}(X,t;W)]+\Delta_r[X_{\mathrm{cl},r}(X,t;W),t^\prime]\right\}\,\mathrm{d}t^\prime+Q_r(X,t;W), \nonumber \\
\eta_{w,v;r}(Z,X,t;W)={}&\frac{4\left[1-\gamma_r(X,t;W)\right](w+2v)\mathcal{Z}_r(Z,X;W)\Delta t_{\mathrm{g},r}(X;W)\hbar k^2}{M} \nonumber \\
{}&-\frac{2\gamma_r(X,t;W)(w+2v)^2[\Delta t_{\mathrm{g},r}(X;W)-(t-t_\mathrm{p})]\hbar k^2}{M}, \label{etaDefinition} \\
\gamma_r(X,t;W)={}&\frac{t-t_\mathrm{p}}{\Delta t_{\mathrm{g},r}(X;W)}, \label{GammaFactor} \\
\Delta t_{\mathrm{g},r}(X;W)={}&\frac{M}{\hbar}\int_{x_\mathrm{g}}^X\frac{\mathrm{d}X^\prime}{K_r(X^\prime;W)}, \nonumber \\
\mathcal{Z}_r(Z,X;W)={}&\frac{ZM}{2\hbar\Delta t_{\mathrm{g},r}(X;W)k}, \nonumber \\
\mathcal{Y}_r(Y,X;W)={}&\frac{YM}{2\hbar\Delta t_{\mathrm{g},r}(X;W)k}, \nonumber \\
X_{\mathrm{cl},r}(X,t;W)={}&X-\frac{\hbar}{M}\int_{t_\mathrm{p}}^tK_r\left[X_{\mathrm{cl},r}(X,t^{\prime\prime};W);W\right]\,\mathrm{d}t^{\prime\prime}, \nonumber \\
R_r(X,t;W)={}&\frac{\mathcal{E}_y\mathcal{E}_x}{2\hbar}\int\chi_r\left[X_{\mathrm{cl},r}(X,t;W),t^\prime\right]\,\mathrm{d}t^\prime, \nonumber \\
Q_r(X,t;W)={}&\frac{\mathcal{E}_y\mathcal{E}_x}{8\hbar}\int\Upsilon^{(1)}_r\left[X_{\mathrm{cl},r}(X,t;W),t^\prime\right]\,\mathrm{d}t^\prime, \nonumber
\end{align}
where $\psi_r^{(0)}(X;W)$ is an amplitude; $\Phi_r(Z,Y,X,t;W)$ is an overall phase; $\eta_{w,v;r}(Z,X,t;W)$ is a phase imperfection, described in more detail below; $\gamma_r(X,t;W)$ is a contraction factor, also described in more detail below; $\Delta t_{\mathrm{g},r}(X;W)$ is the time required classically for the molecule to traverse directly from $x=x_\mathrm{g}$ to $x=X$; $X_{\mathrm{cl},r}(X,t;W)$ is a good approximation to the classical trajectory $X_{\mathrm{cl},r}(X,K_z,K_y,t;W)$; $\mathcal{Z}_r(Z,X;W)$ and $\mathcal{Y}_r(Y,X;W)$ are dimensionless coordinates, normalized by the distance $2\hbar\Delta t_{\mathrm{g},r}(X;W)k/M$ between successive diffraction orders; and $R(X,t;W)$, and $Q(X,t;W)$ are useful shorthands.}

\textcolor{black}{The phase imperfection $\eta_{w,v;r}(Z,X,t;W)$ and contraction factor $\gamma_r(X,t;W)$ arise because the focal point $x_\mathrm{p}$ of the laser pulses is displaced downstream relative to the position $x_\mathrm{g}$ of the mechanical grating. This can be appreciated by noting that the contribution to the wavefunction $\psi^\prime_r(Z,Y,X,t;W)$ for a given value of the composite diffraction order $w+2v$ is optimal at position $Z=2\gamma_r[X,t_\mathrm{p}+\Delta t_{\mathrm{p},r}(X;W);W](w+2v)\hbar\Delta t_{\mathrm{g},r}(X;W)k/M$ ($\mathcal{Z}_r(Z,X;W)=\gamma_r(X,t;W)(w+2v)$) and time $t=t_\mathrm{p}+\Delta t_{\mathrm{p},r}(X;W)$, where
\begin{align}
\Delta t_{\mathrm{p},r}(X;W)={}&\frac{M}{\hbar}\int_{x_\mathrm{p}}^X\frac{\mathrm{d}X^\prime}{K_r(X^\prime;W)} \nonumber
\end{align}
is the time required classically for the molecule to traverse directly from $x=x_\mathrm{p}$ to $x=X$. Setting $Z=2\gamma_r[X,t_\mathrm{p}+\Delta t_{\mathrm{p},r}(X;W);W](w+2v)\hbar\Delta t_{\mathrm{g},r}(X;W)k/M$ and $t=t_\mathrm{p}+\Delta t_{\mathrm{p},r}(X;W)$ in Eqs. \ref{etaDefinition} and \ref{GammaFactor} gives
\begin{align}
{}&\eta_{w,v;r}\left\{\frac{2\gamma_r[X,t_\mathrm{p}+\Delta t_{\mathrm{p},r}(X;W);W](w+2v)\hbar\Delta t_{\mathrm{g},r}(X;W)k}{M},X,t_\mathrm{p}+\Delta t_{\mathrm{p},r}(X;W);W\right\} \nonumber \\
={}&\frac{2\pi\gamma_r[X,t_\mathrm{p}+\Delta t_{\mathrm{p},r}(X;W);W](w+2v)^2(x_\mathrm{p}-x_\mathrm{g})}{x_{\mathrm{T},r}(W)} \label{etaphase} \\
{}&\gamma_r[X,t_\mathrm{p}+\Delta t_{\mathrm{p},r}(X;W);W]=\frac{\Delta t_{\mathrm{p},r}(X;W)}{\Delta t_{\mathrm{g},r}(X;W)}\lesssim1, \label{GammaFactorOptimal} 
\end{align}
where
\begin{align}
x_{\mathrm{T},r}(W)={}&\left[\frac{k^2}{\pi (x_\mathrm{p}-x_\mathrm{g})}\int_{x_\mathrm{g}}^{x_\mathrm{p}}\frac{\mathrm{d}X^\prime}{K_r(X^\prime;W)}\right]^{-1} \label{TalbotLength}
\end{align}
is the Talbot length associated with the mechanical grating to good approximation. Eq. \ref{etaphase} shows that the separation $x_\mathrm{p}-x_\mathrm{g}$ between the laser pulses and the mechanical grating should ideally equal the dilated Talbot length $x_{\mathrm{T},r}(W)/\gamma_r[X,t_\mathrm{p}+\Delta t_{\mathrm{p},r}(X;W);W]$. Eq. \ref{GammaFactorOptimal} shows that $\gamma_r(X,t;W)$ describes a contraction of the laser-pulse diffraction pattern relative to the mechanical-grating diffraction pattern due to the discrepancy in the corresponding propagation times. Eq. \ref{TalbotLength} gives $x_{\mathrm{T},\Omega}(W)=9.43\,\mathrm{mm}$ for a notional molecule with a translational energy of $\hbar W=12.9\,\mathrm{meV}$ ($(2\hbar W/M)^{1/2}=100\,\mathrm{m}\,\mathrm{s}^{-1}$) and perfect orientation of the dipole moment $\pmb{\mu}_0$ along $\mathbf{E}_0(x)$ ($E^{(\mu)}_\Omega(X)-E_\Omega^{(0)}=-f_0(X)\mathcal{E}_z|\pmb{\mu}_0|$).}

\subsection{PECD detection}
\textcolor{black}{Enantiomer-specific molecule detection is realized using PECD. Molecules are ionized in the detection plane located at $x=x_\mathrm{PECD}=1.00\,\mathrm{m}$ at time $t=t_\mathrm{PECD}\sim t_\mathrm{p}+\Delta t_{\mathrm{p},r}(x_\mathrm{PECD};W)$ using an ultraviolet pulse of width $w_\mathrm{PECD}=5.00\,\mathrm{\mu m}$ and duration $\tau_\mathrm{PECD}=100\,\mathrm{fs}$. Enantiomeric excess determination of fenchone down to the level of $1\%$ has already been achieved using this technique \cite{Kastner16a}, demonstrating its feasibility.}

\textcolor{black}{As the ionizing pulse width is smaller than the separation between diffraction orders ($w_\mathrm{PECD}\ll2\hbar\Delta t_{\mathrm{g},r}(x_\mathrm{PECD};W)k/M$) and the molecules move only slightly during the ionization ($\hbar K_{r,x}(x_\mathrm{PECD},K_z,K_y;W)\tau_\mathrm{PECD}/M\ll w_x^\prime,w_x$), we take the probability density for detecting a given isotopologue of a given conformer of a given enantiomer to be
\begin{align}
P(Z)={}&\sum_r\int P_r(Z;W)\,\mathrm{d}W, \label{P(Z)} \nonumber
\end{align}
where 
\begin{align}
P_r(Z;W)\propto{}&\int |\psi^\prime_r(Z,Y, x_\mathrm{PECD},t_\mathrm{PECD};W)|^2\,\mathrm{d}Y w_\mathrm{PECD}\,p_r(W)\nonumber
\end{align}
is the probability density for a given rotational state and translation energy, assuming complete vertical integration.}

\begin{figure}[h!]
\centering
\includegraphics[width=0.5\linewidth]{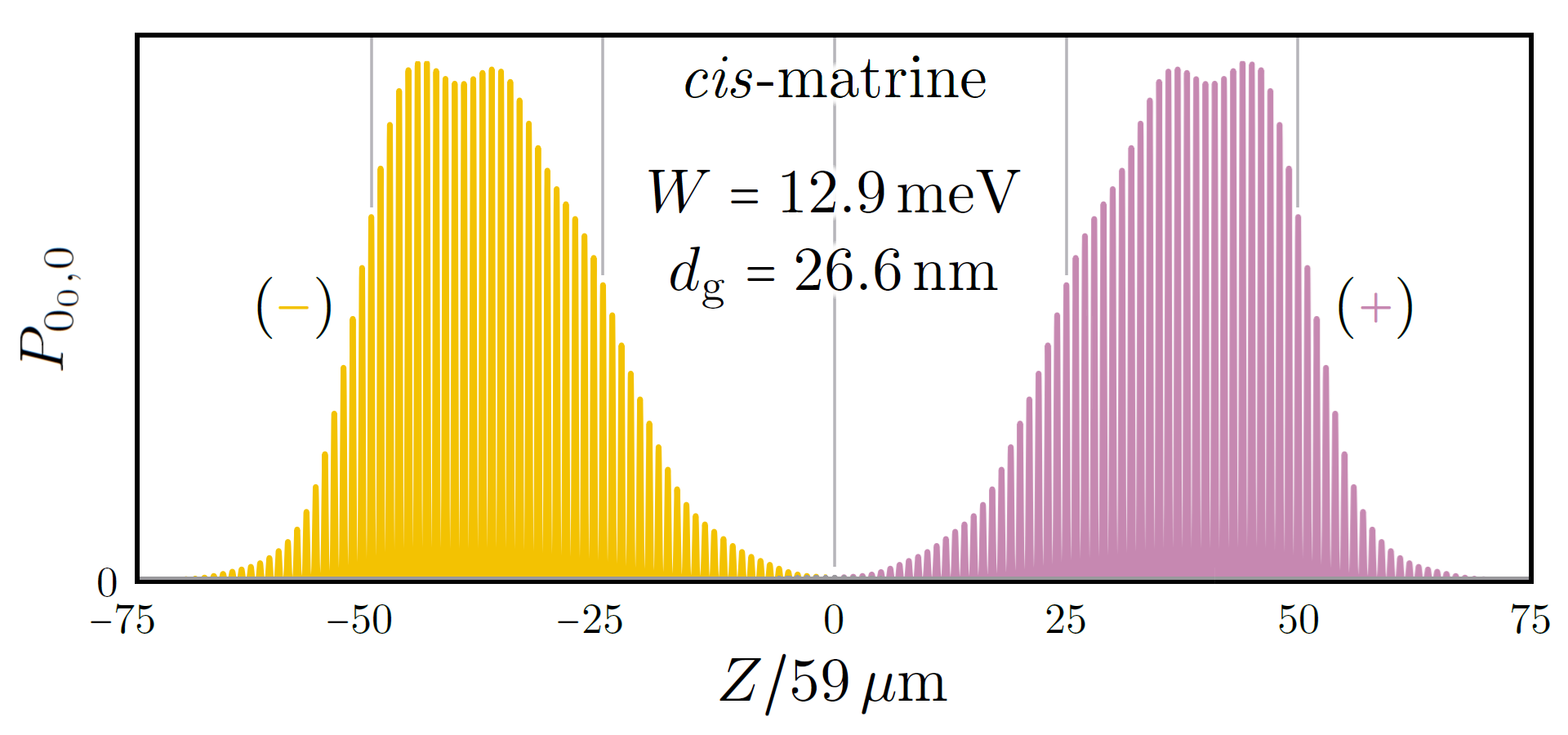}
\caption{\small \textcolor{black}{As in Fig. \ref{Fig3} but for the \textit{cis} conformation.}}
\label{Fig7}
\end{figure}

\textcolor{black}{Shown in Fig. \ref{Fig7} are results for ($\pm$)-\textit{cis}-matrine.}

\end{appendix}
\end{document}